\documentclass[12pt]{cernart}

\usepackage{epsfig}
\usepackage{graphicx} 
\usepackage{amsmath,amssymb}
\usepackage{lineno}
\newcommand{\defmath}[2] {\def#1{\ifmmode{#2}\else\mbox{${#2}$}\fi}}
\newcommand{\defdecay}[2] {\def#1{\ifmmode{#2}\else\mbox{${#2}$}\fi}}

\newcommand{\defunit}[2] {\def#1{\ifmmode\mathrm{#2}\else\mbox{$\mathrm{#2}$}\fi
}} 

\defunit{\mum}{\mu m}
\defunit{\mus}{\mu s}
\defunit{\taus}{\tau_S}
\defunit{\degrees}{^{\circ}}
\defunit{\xo}{X_0}

\defdecay{\k} {K}
\defdecay{\kz} {K^0}
\defdecay{\kbar} {\overline{K^{0}}}
\defdecay{\cascadebar} {\overline{\Xi^0}}
\defdecay{\lambdabar} {\overline{\Lambda}}
\defdecay{\kone} {K_1}
\defdecay{\ktwo} {K_2}
\defdecay{\ks} {K_S}
\defdecay{\kl} {K_L}
\defdecay{\ksl} {K_{S,L}}
\defdecay{\KL} {\kl}
\defdecay{\KS} {\ks}
\defdecay{\sbar} {\overline{\mathrm{s}}}
\defdecay{\dbar} {\overline{\mathrm{d}}}

\defdecay{\pipi} {\pi^{+}\pi^{-}}
\defdecay{\twopi} {2\pi}
\defdecay{\kpipi} {\k \rightarrow \pipi}
\defdecay{\ktwopi} {\ks \rightarrow \twopi}
\defdecay{\kspipi} {\ks \rightarrow \pipi}
\defdecay{\kstwopi} {\kl \rightarrow \twopi}
\defdecay{\klpipi} {\kl \rightarrow \pipi}
\defdecay{\kltwopi} {\k \rightarrow \twopi}
\defdecay{\pipipi} {\pi\pi\pi}
\defdecay{\threepi} {3\pi}

\defdecay{\pz}{\pi^0}
\defdecay{\pio} {\pi^0}
\defdecay{\twopio} {2 \pi^0}
\defdecay{\piopio} {\pi^0 \pi^0}
\defdecay{\pipin} {\piopio}
\defdecay{\ktwopio} {\k \rightarrow \twopio}
\defdecay{\kpiopio} {\k \rightarrow \piopio}
\defdecay{\kltwopio} {\kl \rightarrow \twopio}
\defdecay{\klpiopio} {\kl \rightarrow \piopio}
\defdecay{\kstwopio} {\ks \rightarrow \twopio}
\defdecay{\kspiopio} {\ks \rightarrow \piopio}
\defdecay{\klthreepich} {\kl \rightarrow \pio\pi^{+}\pi^{-}}

\defdecay{\kspimumu} {\ks \rightarrow \pio \mu^+ \mu^-}
\defdecay{\klpimumu} {\kl \rightarrow \pio \mu^+ \mu^-}

\defdecay{\kspiee} {\ks \rightarrow \pio e^+ e^-}
\defdecay{\klpiee} {\kl \rightarrow \pio e^+ e^-}
\defdecay{\kspipiD} {\ks \rightarrow \pio \pioD }
\defdecay{\kspipid} {\ks \rightarrow \pio \piod }
\defdecay{\kspiDpiD} {\ks \rightarrow \pioD \pioD }
\defdecay{\kspidpid} {\ks \rightarrow \piod \piod }
\defdecay{\kspipidd} {\ks \rightarrow \pio \piodd}
\defdecay{\kpiee} {K \rightarrow \pi e e}
\defdecay{\piee} {\pio e^+ e^-}
\defdecay{\pipid} {\pio \piod}
\defdecay{\pidpid} {\piod \piod}
\defdecay{\pipidd} {\pio \piodd}
\defdecay{\p0pipi} {\pio \pi^{+} \pi^{-} }
\defdecay{\kleegg} {\kl\rightarrow e e\gamma\gamma}
\defdecay{\klmumugg} {\kl\rightarrow \mu\mu\gamma\gamma}
\defdecay{\klpigg} {\kl\rightarrow \pio \gamma \gamma}
\defdecay{\kseegg} {\ks\rightarrow e e\gamma\gamma}
\defdecay{\eegg} {e e\gamma\gamma}
\defdecay{\kleeg} {\kl\rightarrow e e\gamma}
\defdecay{\klp0pipi} {\kl\rightarrow \pio \pi^+ \pi^- }
\defdecay{\BRkspiee} {B ( \ks \rightarrow \pio e^+ e^- ) }
\defdecay{\BRklpiee} {B ( \kl \rightarrow \pio e^+ e^- ) }
\defdecay{\BRkspipid} {B ( \kspipid ) }
\defdecay{\BRkspidpid} {B ( \kspidpid ) }
\defdecay{\BRklp0pipi} {B ( \klp0pipi) }
\defdecay{\BRp0pipi} {B ( \p0pipi) }
\defdecay{\Xilampi} {\Xi^0 \rightarrow \Lambda \pio}
\defdecay{\lamppi} {\Lambda \rightarrow p \pi^{-}}
\defdecay{\Xilamppipio} {\Xi^{0}\rightarrow\Lambda (p \pi^{-})\pio }
\defdecay{\lampev} {\Lambda \rightarrow p e^{-} \nu}
\defdecay{\Xilampevpio} {\Xi^{0}\rightarrow\Lambda (p e^{-} \nu)\pio }
\defdecay{\Xisigmaenv} {\Xi^0 \rightarrow \Sigma^{+} e^{-} \nu}
\defdecay{\sigppio} {\Sigma^{+} \rightarrow  p \pio}
\defdecay{\Xisigmappioenv} {\Xi^0 \rightarrow \Sigma^{+}(p \pio) e^{-} \nu}
\defdecay{\AXilampi} {\overline{\Xi^0}\rightarrow \overline{\Lambda} \pio  }
\defdecay{\Alamppi} {\overline{\Lambda} \rightarrow \overline{p} \pi^{+}}
\defdecay{\Alampev} {\overline{\Lambda} \rightarrow \overline{p} e^{+} \, \overline{\nu}}
\defdecay{\AXisigmaenv} {\overline{\Xi^0} \rightarrow \overline{\Sigma^{+}} e^{+} \nu}
\defdecay{\Asigppio} {\overline{\Sigma^{+}} \rightarrow  \overline{p} \pio}
\defdecay{\klkef} {\kl\rightarrow \pio \pi^{\pm} e^{\mp} \nu }
\defdecay{\kef} {\pio \pi^{\pm} e^{\mp} \nu }
\defdecay{\klketh} {\kl\rightarrow \pi^{\pm} e^{\mp} \nu }
\defdecay{\keth} {\pi^{\pm} e^{\mp} \nu }

\defdecay{\dalconv} {{\bf\boldmath \pio}(\gamma\gamma){\bf\boldmath \piod}(\gamma_{conv}(ee\hspace {-3.5mm}\nearrow)ee\hspace {-3.5mm}\nearrow)}
\defdecay{\ndalconv} {{\bf\boldmath \pio}(\gamma_{conv}(ee\hspace {-3.5mm}\nearrow)\gamma){\bf\boldmath \piod}(\gamma ee\hspace {-3.5mm}\nearrow)}
\defdecay{\dalcomp} {{\bf\boldmath \pio}(\gamma\gamma){\bf \boldmath\piod}(\gamma_{comp}(e^{-})e^{+}e^{-}\hspace {-1.5mm}\nearrow)}

\defdecay{\pipiconv} {{\bf\boldmath \pio}(\gamma\gamma){\bf\boldmath \pio}(\gamma_{conv}(ee\hspace {-6.5mm}\nearrow)\gamma_{conv}(ee\hspace {-6.5mm}\nearrow))}

\newcommand{\klpimunu}{\kl \to \pi^\pm \mu^\mp \, \nu_\mu}
\defdecay{\klpipipic} {\kl \rightarrow \pio \pi^+ \pi^-}
\defdecay{\kspipic} {\ks \rightarrow \pi^+ \pi^-}
\defdecay{\kspiee} {\ks \rightarrow \pio e^+ e^-}
\defdecay{\kspimumu} {\ks \rightarrow \pio \mu^+ \mu^-}
\defdecay{\klmumugg} {\kl \rightarrow \mu^+ \mu^- \gamma \gamma}
\defdecay{\klpiogg} {\kl \rightarrow \pio \gamma \gamma}
\defdecay{\kspipig} {\ks \rightarrow \pi^+ \pi^- \gamma}
\defdecay{\klmufour} {\kl \rightarrow \pio \pi^{\pm} \mu^{\mp} \nu}
\defdecay{\klpimunug} {\kl \rightarrow \pi^{\pm} \mu^{\mp} \nu \gamma}

\defdecay{\pip} {\pi^+}
\defdecay{\pim} {\pi^-}
\defdecay{\twopic} {\pi^+ \pi^-}
\defdecay{\pipic} {\twopic}
\defdecay{\ktwopic} {\k \rightarrow \twopic}
\defdecay{\kltwopic} {\kl \rightarrow \twopic}
\defdecay{\kstwopic} {\ks \rightarrow \twopic}

\defdecay{\piod} {\pi_{D}^0}
\defdecay{\pioD} {\pi_{Dalitz}^0}
\defdecay{\piodd} {\pi_{DD}^0}
\defdecay{\piopiod} {\pi^0 \pi_{D}^0}
\defdecay{\eeg} {ee\gamma}
\defdecay{\pioeeg} {\pi^0 \rightarrow \eeg}
\defdecay{\kpiopiod} {\k \rightarrow \piopiod}

\defdecay{\kethree} {\mathrm{K_{e3}}}
\defdecay{\pienu} {\pi e \nu}
\defdecay{\klethree} {\kl \rightarrow \pienu}
\defdecay{\kmuthree} {\mathrm{K_{\mu 3}}}
\defdecay{\pimunu} {\pi \mu \nu}
\defdecay{\klmuthree} {\kl \rightarrow \pimunu}
\defdecay{\threepio} {3\pi^0}
\defdecay{\klthreepio} {\kl \rightarrow \threepio}

\defdecay{\Lam}{\Lambda}
\defdecay{\Lambar}{\bar{\Lambda}}
\defdecay{\etagg}{\eta \rightarrow \gamma \gamma}
\defdecay{\etathreepio} {\eta \rightarrow \threepio}
\defdecay{\piogg} {\pi^0 \rightarrow \gamma \gamma}
\defdecay{\meegg} {m_{ee \gamma \gamma}}
\defdecay{\meeggg} {m_{ee \gamma \gamma \gamma}}
\defdecay{\mgg} {m_{\gamma \gamma}}
\defdecay{\mmumugg} {m_{\mu\mu \gamma \gamma}}

\defmath{\dvertex}{d_{vertex}}
\defmath{\mgg}{m_{\gamma\gamma}}
\defmath{\chisq}{\chi^2}
\defmath{\rel}{\chi^2}
\defmath{\mone}{m_1}
\defmath{\mtwo}{m_2}

\defmath{\pk}{p_K}
\defmath{\pt}{p_T}
\defmath{\ptp}{{p_T}'}
\defmath{\ptpsq}{{p'_T}^2}
\defmath{\mpp}{m_{\pi\pi}}

\defmath{\asl} {\alpha_{SL}}
\defmath{\asloo} {\alpha^{00}_{SL}}
\defmath{\aslpm} {\alpha^{+-}_{SL}}
\defmath{\Dasl} {\Delta \alpha_{SL}}

\defmath{\als} {\alpha_{LS}}
\defmath{\alsoo} {\alpha^{00}_{LS}}
\defmath{\alspm} {\alpha^{+-}_{LS}}
\defmath{\Dals} {\Delta \alpha_{LS}}

\defmath{\btag} {\beta_{tag}}
\defmath{\btagoo} {\beta^{00}_{tag}}
\defmath{\btagpm} {\beta^{+-}_{tag}}
\defmath{\Dbtag} {\Delta \beta_{tag}}

\defmath{\wpm} {W^{+-}}
\defmath{\woo} {W^{00}}
\defmath{\wooo} {W^{000}}
\defmath{\Dw} {\Delta W}

\defmath{\wt}{W(\tau)}
\defmath{\Dp}{\mathrm{D_p}}

\defmath{\etas}{\eta_S}
\defmath{\etal}{\eta_L}
\defmath{\etasl}{\eta_{S,L}}
\defmath{\lamc}{\lambda^{+-}}
\defmath{\lamn}{\lambda^{00}}
\defmath{\DRint}{(\DR)_{\mbox{\scriptsize intensity}}}
\defmath{\DRgeom}{(\DR)_{\mbox{\scriptsize geometry}}}

\defmath{\R}{R}
\defmath{\DR}{\Delta \R}
\defmath{\epp}{\varepsilon^{\prime}}
\defmath{\vep}{\varepsilon}
\defmath{\epe}{\epp/\vep}
\defmath{\Ree}{\mathcal{R}\!\mathit{e}(\eprime/\epsilon)}
\defmath{\epm}{\eta_{+-}}
\defmath{\eoo}{\eta_{00}}

\defmath{\mum}{\mu\mathrm{m}}
\defmath{\mus}{\mu\mathrm{s}}
\defmath{\degrees}{^{\circ}}
\defmath{\taus}{\tau_S}
\defmath{\taul}{\tau_L}

\defmath{\about}{\sim}
\defmath{\eop}{E/p}
\defmath{\Qx} {Q_x}
\defmath{\twotrack}{2track}
\defmath{\etot}{E_{tot}}
\defmath{\mk}{m_K}
\defmath{\stat}{\mbox{stat}}
\defmath{\syst}{\mbox{syst}}
\defmath{\rcog}{R_{cog}}
\defmath{\mm}{m}
\defmath{\mee}{m_{ee}}
\defmath{\mpi}{m_{\pio}}

\defdecay{\pimumu} {\pio \mu^+ \mu^-}

\usepackage{amssymb}

\begin{document}
\begin{titlepage}
\docnum{CERN-PH-EP/2004-025}
\date{15~June~2004}
\title{\bf \Large Observation of the rare decay {\boldmath $\kspimumu$}}

\begin{Authlist}
\begin{center}
\  \\[0.2cm]
 %
%
J.R.~Batley,
G.E.~Kalmus\footnotemark[1],
C.~Lazzeroni,
D.J.~Munday,
M.~Patel,
M.W.~Slater,
S.A.~Wotton \\
{\em \small Cavendish Laboratory, University of Cambridge, Cambridge, CB3 0HE,
U.K.\footnotemark[2]} \\[0.2cm]
 R.~Arcidiacono,
 G.~Bocquet,
 A.~Ceccucci,
 D.~Cundy\footnotemark[3],
 N.~Doble\footnotemark[4],
 V.~Falaleev,
 L.~Gatignon,
 A.~Gonidec,
 P.~Grafstr\"om,
 W.~Kubischta,
F.~Marchetto\footnotemark[5],
 I.~Mikulec\footnotemark[6],
 A.~Norton,
 B.~Panzer-Steindel,
P.~Rubin\footnotemark[7],
 H.~Wahl\footnotemark[8] \\
{\em \small CERN, CH-1211 Gen\`eve 23, Switzerland} \\[0.2cm]
E.~Monnier\footnotemark[9],
E.~Swallow,
R.~Winston\\
{\em \small The Enrico fermi Institute, The University of Chicago, Chicago, Illinois, 60126, U.S.A.}\\[0.2cm]
E.~Goudzovski,
D.~Gurev,
P.~Hristov\footnotemark[10],
V.~Kekelidze,
V.~Kozhuharov,
L.~Litov,
D.~Madigozhin,
N.~Molokanova,
Yu.~Potrebenikov,
S.~Stoynev,
A.~Zinchenko\\
{\em \small Joint Institute for Nuclear Research, Dubna, Russian    Federation} \\[0.2cm]
 R.~Sacco\footnotemark[11],
 A.~Walker \\
{\em \small Department of Physics and Astronomy, University of    Edinburgh, JCMB King's Buildings, Mayfield Road, Edinburgh,    EH9 3JZ, U.K.} \\[0.2cm]
%
W.~Baldini,
P.~Dalpiaz,
J.~Duclos,
P.L.~Frabetti,
A.~Gianoli,
M.~Martini,
F.~Petrucci,
M.~Scarpa,
M.~Savri\'e \\
{\em \small Dipartimento di Fisica dell'Universit\`a e Sezione    dell'INFN di Ferrara, I-44100 Ferrara, Italy} \\[0.2cm]
%
A.~Bizzeti\footnotemark[12],
M.~Calvetti,
G.~Graziani,
E.~Iacopini,
M.~Lenti,
F.~Martelli\footnotemark[13],
G.~Ruggiero,
M.~Veltri\footnotemark[13] \\
{\em \small Dipartimento di Fisica dell'Universit\`a e Sezione    dell'INFN di Firenze, I-50125 Firenze, Italy} \\[0.2cm]
%
%
M.~Behler,
K.~Eppard,
 M.~Eppard,
 A.~Hirstius,
 K.~Kleinknecht,
 U.~Koch,
L.~Masetti,
P.~Marouelli,
U.~Moosbrugger,
C.~Morales Morales,
 A.~Peters,
 R.~Wanke,
 A.~Winhart \\
{\em \small Institut f\"ur Physik, Universit\"at Mainz, D-55099 Mainz,
Germany\footnotemark[14]} \\[0.2cm]
A.~Dabrowski,
T.~Fonseca Martin,
M.~Szleper,
M.~Velasco \\
{\em \small Department of Physics and Astronomy, Northwestern University, Evanston Illinois 60208-3112, U.S.A.}
 \\[0.2cm]
G.~Anzivino,
P.~Cenci,
E.~Imbergamo,
G.~Lamanna,
P.~Lubrano,
A.~Michetti,
A.~Nappi,
M.~Pepe,
M.C.~Petrucci,
M.~Piccini,
M.~Valdata \\
{\em \small Dipartimento di Fisica dell'Universit\`a e Sezione    dell'INFN di Perugia, I-06100 Perugia, Italy} \\[0.2cm]
%
%
 C.~Cerri,
G.~Collazuol,
F.~Costantini,
 R.~Fantechi,
 L.~Fiorini,
 S.~Giudici,
 I.~Mannelli,
G.~Pierazzini,
 M.~Sozzi \\
{\em \small Dipartimento di Fisica, Scuola Normale Superiore e Sezione dell'INFN di Pisa, I-56100 Pisa, Italy} \\[0.2cm]
%

 C.~Cheshkov\footnotemark[10],
 J.B.~Cheze,
 M.~De Beer,
 P.~Debu,
 G.~Gouge,
 G.~Marel,
 E.~Mazzucato,
 B.~Peyaud,
 B.~Vallage \\
{\em \small DSM/DAPNIA - CEA Saclay, F-91191 Gif-sur-Yvette, France} \\[0.2cm]
M.~Holder,
 A.~Maier,
 M.~Ziolkowski \\
{\em \small Fachbereich Physik, Universit\"at Siegen, D-57068 Siegen,
Germany\footnotemark[15]} \\[0.2cm]
C.~Biino,
N.~Cartiglia,
M.~Clemencic,
S.~Goy-Lopez,
E.~Menichetti,
N.~Pastrone \\
{\em \small Dipartimento di Fisica Sperimentale dell'Universit\`a e    Sezione dell'INFN di Torino,  I-10125 Torino, Italy} \\[0.2cm]
 W.~Wislicki \\
{\em \small Soltan Institute for Nuclear Studies, Laboratory for High    Energy
Physics,  PL-00-681 Warsaw, Poland\footnotemark[16]} \\[0.2cm]
H.~Dibon,
M.~Jeitler,
M.~Markytan,
G.~Neuhofer,
L.~Widhalm \\
{\em \small \"Osterreichische Akademie der Wissenschaften, Institut  f\"ur
Hochenergiephysik,  A-10560 Wien, Austria\footnotemark[17]} \\[1cm]
\vspace{0.5cm}
\it{Submitted for publication in Physics Letters B.}
\rm
\end{center}
\setcounter{footnote}{0}
\footnotetext[1]{Present address: Rutherford Appleton Laboratory,
Chilton, Didcot, OX11 0QX, UK}
\footnotetext[2]{ Funded by the U.K.    Particle Physics and Astronomy Research Council}
\footnotetext[3]{Present address: Istituto di Cosmogeofisica del CNR di Torino, I-10133 Torino, Italy}
\footnotetext[4]{Also at Dipartimento di Fisica dell'Universit\`a e Sezione dell'INFN di Pisa, I-56100 Pisa, Italy}
\footnotetext[5]{On leave from Sezione dell'INFN di Torino,  I-10125 Torino, Italy}
\footnotetext[6]{ On leave from \"Osterreichische Akademie der Wissenschaften, Institut  f\"ur Hochenergiephysik,  A-1050 Wien, Austria}
\footnotetext[7]{On leave from University of Richmond, Richmond, VA, 23173,
USA; supported in part by the US NSF under award \#0140230}
\footnotetext[8]{Also at Dipartimento di Fisica dell'Universit\`a e Sezione dell'INFN di Ferrara, I-44100 Ferrara, Italy}
\footnotetext[9]{Also at Centre de Physique des Particules de Marseille, IN2P3-CNRS, Universit\'e
de la M\'editerran\'ee, Marseille, France}
\footnotetext[10]{Present address CERN, CH-1211 Gen\`eve 23, Switzerland}
\footnotetext[11]{Present address Laboratoire de l'Acc\'elerateur Lin\'eaire, IN2P3-CNRS, Universit\'e de Paris-Sud, 91898 Orsay, France}
\footnotetext[12]{ Dipartimento di Fisica dell'Universit\`a di Modena e Reggio Emilia, via G. Campi 213/A I-41100, Modena, Italy}
\footnotetext[13]{ Istituto di Fisica, Universit\`a di Urbino, I-61029  Urbino, Italy}
\footnotetext[14]{ Funded by the German Federal Minister for    Research and Technology (BMBF) under contract 7MZ18P(4)-TP2}
\footnotetext[15]{ Funded by the German Federal Minister for Research and Technology (BMBF) under contract 056SI74}
\footnotetext[16]{Supported by the Committee for Scientific Research grants
5P03B10120, SPUB-M/CERN/P03/DZ210/2000 and SPB/CERN/P03/DZ146/2002}
\footnotetext[17]{Funded by the Austrian Ministry for Traffic and
Research under the
contract GZ 616.360/2-IV GZ 616.363/2-VIII,
and by the Fonds f\"ur   Wissenschaft und Forschung FWF Nr.~P08929-PHY}

\end{Authlist}

\end{titlepage}

\setcounter{footnote}{0}


\begin{abstract}

A search for the decay $\kspimumu$ has been made 
by the NA48/1 Collaboration at the CERN SPS accelerator. 
The data were collected during 2002 with a high-intensity $\ks$ beam.  
Six events were found with a background expectation of $0.22^{+0.18}_{-0.11}$ 
events.  Using a vector matrix element and unit form factor,
the measured branching ratio is
{ $$  B(\kspimumu) = [2.9^{+1.5}_{-1.2}({\it stat})\pm {0.2}({\it syst})] \times 10^{-9}. $$}

\end{abstract}


\vspace{0.2cm}

\section{Introduction}
\label{sec:Introduction}
 
This paper reports the first observation of the decay
$\kspimumu$ and a measurement of its branching ratio.
The analysis was carried out on data taken during 2002
by the NA48/1 experiment at the CERN SPS,
which also recently reported the first observation of the decay
$\kspiee$~\cite{bib:piee}.

The physics interest of the $\kspimumu$ decay
is that it measures the indirect CP violating contribution of the decay
$\klpimumu$,
thereby allowing the direct CP violating component of the $\KL$ decay to be extracted.
This can provide input to the determination of the imaginary part
of the element $V_{td}$ of the Cabibbo, Kobayashi, Maskawa (CKM)
mixing matrix.
In addition, the decay $\kspimumu$
can be used to study the structure of the $K\rightarrow\pi\gamma^*$
form factor.
This is particularly interesting if combined with the
$\kspiee$ results since
both decays are expected to be dominated by the exchange of a single
virtual photon ($K \rightarrow \pi \gamma^* \rightarrow \pi\ell^+\ell^-$).

\section{Experimental Setup}

The beam line and detector built by the NA48 Collaboration 
to measure the $Re(\epsilon^{\prime}/ \epsilon)$ parameter 
\cite{bib:epsi} from a comparison of $K_{S,L}\to\pi^+\pi^-,\pi^0\pi^0$ decays was used,
with the modifications described below.
\subsection{Beam}

The experiment was performed at the CERN SPS accelerator and used a
400\,GeV proton beam impinging on a Be target to produce a neutral beam.
The spill length was 4.8\,s out of a 16.8\,s cycle time.  The proton intensity 
was fairly constant during the spill with a mean of $5 \times 10^{10}$  
protons per spill. 

The neutral kaon beam line was modified as follows:\,
the \kl\ beam was blocked
and a small, additional sweeping magnet was installed above
the \ks\ collimator.  In order to reduce the number of photons in 
the neutral beam, primarily from $\pi^0$ decays, a platinum absorber 24\,mm 
thick was placed in the beam between the target and the main
sweeping magnet which deflected charged particles into the collimator.  
The 5.1\,m long \KS\ collimator,  with its axis
at an angle of 4.2\,mrad to the proton beam direction, selected a beam of 
neutral long-lived particles (\KS, \KL, $\Lambda^0$, $\Xi^0$, $n$, $\gamma$'s, etc.). 
On average, there were  $2 \times 10^{5}$ \KS\, decays per spill  in
the fiducial volume downstream of the collimator, with 
momenta between 60 and 200\,GeV
(and a mean energy of $\sim 110$\,GeV).
An average of 0.017 \KL\ decays were expected for every \KS\ decay 
within the first three \KS\ lifetimes from the end
of the collimator.

\subsection{Detector}

In order to minimize interactions of the neutral beam,  
the collimator was immediately followed by a 90\,m long evacuated tank.
The tank was terminated by a 0.3\% radiation length\,($X_0$) thick Kevlar window,
except in a region close to the beam
which continued in a vacuum pipe through the centre of the downstream detectors.

\subsubsection{Tracking}

The tracking was performed with a spectrometer housed in a helium gas volume. It  consisted of two 
drift chambers before and two after a dipole magnet with a horizontal 
transverse momentum kick of 265\,MeV. Each chamber had four 
views,
each of which had two sense wire planes.  The resulting space points were 
typically reconstructed with a resolution of 150\,$\mu$m in each 
projection.  The spectrometer momentum resolution was parameterised as:
\begin{equation*}
\nonumber
\sigma_p/p = 0.48 \% \oplus 0.015\% \cdot p,
\end{equation*}
\noindent where $p$ is in GeV. This gave a resolution of 
3\,MeV on the reconstructed kaon mass in \kspipi\ decays.
The track time resolution was 1.4\,ns.

\subsubsection{Electromagnetic Calorimetry}

The detection and measurement of electromagnetic showers 
were performed with a 27\,$X_0$ deep
liquid krypton calorimeter (LKr).
The energy resolution
was parameterised as \cite{bib:unal}:
\begin{equation*}
\nonumber
\sigma(E) / E = 3.2\% /\sqrt{E} \oplus 9 \% / E \oplus 0.42 \%,
\end{equation*}
where $E$ is in GeV.
The calorimeter was subdivided into 13,500 cells
of transverse dimension 2\,cm $\times $ 2\,cm,
which resulted in a transverse position resolution
better than 1.3 mm for single photons with energy above 20\,GeV. 
The $\pi^0$ mass resolution was
0.8\,MeV, while 
the time resolution of the calorimeter for a single shower was
better than 300\,ps.

\subsubsection{Scintillator Detectors and Muon Detector}

A scintillator hodoscope was located between the
spectrometer and the LKr. It consisted of two planes, segmented
in horizontal and vertical strips respectively,  
with each plane arranged in four quadrants.
The time resolution for the hodoscope system was 200\,ps.
Downstream of the LKr calorimeter was an iron-scintillator sandwich 
hadron calorimeter\,(HCAL), followed by muon counters\,(MUC)  which 
consisted of three planes 
of plastic scintillators, each shielded by an 80\,cm thick iron wall.  
The first two planes, $M1X$ and $M1Y$,
consisted of 25\,cm wide horizontal ($M1X$) and vertical ($M1Y$) scintillator
strips, with a length of 2.7\,m.  The third plane, $M2X$, consisted of
horizontal strips of width 44.6\,cm and was mainly used to measure 
the efficiency of the $M1X$ and $M1Y$ counters.
The central strip in  each plane was split with a gap of 21\,cm
to accommodate the beam pipe.
The fiducial volume of the experiment was principally determined by the 
LKr calorimeter acceptance, together with 
seven rings of scintillation counters
which surrounded the decay volume to veto activity outside this region.

\subsubsection{Trigger and Readout}
\label{sec:t_and_r}

The detector was sampled every 25\,ns, and samples were 
recorded in  time windows of $200$\,ns. 
The extended time window 
allowed the rate of accidental activity to be investigated 
from the  time sidebands.

The trigger selection for \kspimumu\ candidates
consisted of a first-stage hardware trigger
followed by a second-stage software trigger.
The hardware trigger, for the first 40\% of the data taken, selected events 
satisfying the following conditions:
  \begin{itemize}
  \item at least one hit in each of the horizontal and vertical planes
        of the hodoscope, within the same quadrant;
  \item at least three hit wires in at least three views of DCH1 integrated over 200\,ns;
  \item a track vertex located within 90\,m of the end of the collimator;
  \item no hit in the two ring scintillator counters farthest downstream;
  \item a two-muon signal ($2\mu_{\mathrm{tight}}$) from the muon counters.  
  \end{itemize}
The $2\mu_{\mathrm{tight}}$ signal required at least two hits in each of the
first two muon counter planes ($M1X$ and $M1Y$).

For the last 60\% of the data taken,
a second, parallel, hardware trigger component was added in which the 
$2\mu_{\mathrm{tight}}$ condition was replaced by the following requirements:
  \begin{itemize}
  \item hadron calorimeter energy less than 10\,GeV;
  \item electromagnetic calorimeter energy greater than 15\,GeV;
  \item a two-muon signal ($2\mu_{\mathrm{loose}}$) from the muon counters. 
  \end{itemize}
The $2\mu_{\mathrm{loose}}$ signal was similar to the $2\mu_{\mathrm{tight}}$ signal
but allowed one of the first two muon planes ($M1X$ or $M1Y$) to contain
only a single hit.
The addition of the $2\mu_{\mathrm{loose}}$ trigger significantly improved the
trigger acceptance for the $\kspimumu$ signal,
as discussed below in Section~\ref{sec:trig_effy}.

The software trigger (for the entire data set) required:
\begin{itemize} 
 \item at least two
 tracks in the drift chambers that were not associated with 
energetic\,($>$5\,GeV)
clusters in the LKr;
  \item less than 10\,GeV in the hadronic calorimeter;
  \item at least one hit in the first two muon counter planes.
\end{itemize}
Events that satisfied the trigger conditions were recorded and
reprocessed with improved calibrations to obtain the final data sample.

\section{Data Analysis}
The signal channel \kspimumu\ required the identification of
two muons of opposite charge accompanied by two additional clusters
(photons) in the LKr.
The kaon energy was estimated from the sum of the track momenta
and cluster energies
and was required to be between 60 and 200\,GeV.

The invariant mass, $m_{\mu \mu}$,
for the two muons from a $\kspimumu$ decay is limited by kinematics to:
\begin{equation*}
\nonumber
0.211\,{\rm GeV} < m_{\mu\mu} < 0.363\,{\rm GeV}
\end{equation*}
where the lower limit corresponds to $m_{\mu}+m_{\mu}$, while the upper 
limit is given by $m_K-m_{\pi^0}$.

The known \KS\, mass was used as a constraint\,(see Eq.~\ref{eq:zk} below)
to reconstruct the longitudinal vertex position using information
from the charged tracks and the clusters.
The invariant mass, $\mgg$, of the two photons
was reconstructed using the measured positions and energies of the two clusters,
assuming this vertex position.
The total invariant mass, $m_{\mu\mu\pi}$,
was calculated using the vertex reconstructed from the charged tracks alone,
with the known $\pi^0$ mass imposed as a constraint
to improve the kaon mass resolution.

\subsection{Signal and Control regions}

The similarity between the kinematics of the 
\klp0pipi\ and \kspimumu\
channels allowed the former channel to be used 
to evaluate the resolutions
$\sigma_{\mgg}$ and $\sigma_{m_{\mu\mu\pi}}$
on ${\mgg}$ and
$m_{\mu\mu\pi}$, respectively.
These were measured to be $\sigma_{\mgg}=0.8$\,MeV and 
$\sigma_{m_{\mu\mu\pi}}=3$\,MeV,
in agreement with a detailed  Monte Carlo simulation based on GEANT~\cite{bib:geant}.

A signal region and a control region were defined in the
$(m_{\gamma\gamma}, m_{\mu\mu\pi})$ plane:
\begin{itemize}
\item {\it SIGNAL REGION:}\\
  $| m_{\gamma \gamma} - m_{\pi^0}| \le 2.5 \,
  \sigma_{\mgg}$ and $| m_{\mu\mu\pi} - m_K| \le 2.5 \, \sigma_{m_{\mu\mu\pi}}$;
\item {\it CONTROL REGION:}\\
  $3 \,\sigma_{\mgg} \le | m_{\gamma \gamma} - m_{\pi^0}| \le 6 \, \sigma_{\mgg}$
  and
  $3 \, \sigma_{m_{\mu\mu\pi}}\le | m_{\mu\mu\pi} - m_K| \le 6 \, \sigma_{m_{\mu\mu\pi}}$.
\end{itemize}
The signal and control regions were kept masked until the 
analysis cuts were finalised in order to keep the event selection unbiased. 
The cuts to reject the background were set using both data and Monte Carlo simulation.
An extended control region was used to estimate the  
contributions from various backgrounds.

\subsection{Event selection}

In order to identify muons,
charged tracks reconstructed in the spectrometer were
extrapolated to the MUC planes
and associated with MUC hits using cuts on the spatial separation
and time difference between the extrapolated track and the MUC hit.
Multiple scattering was taken into account before applying the spatial cut,
and light propagation along the MUC strips was taken into account
before applying the time-difference cut.
For muons with momentum greater than 10\,GeV,
the efficiency of the MUC counters was found to be $0.99 \pm 0.01$.

\begin{figure}[hbtp]
  \vspace{9pt}
  \centerline{\hbox{ \hspace{0.0in}
    \epsfxsize=3.0in
  \epsffile{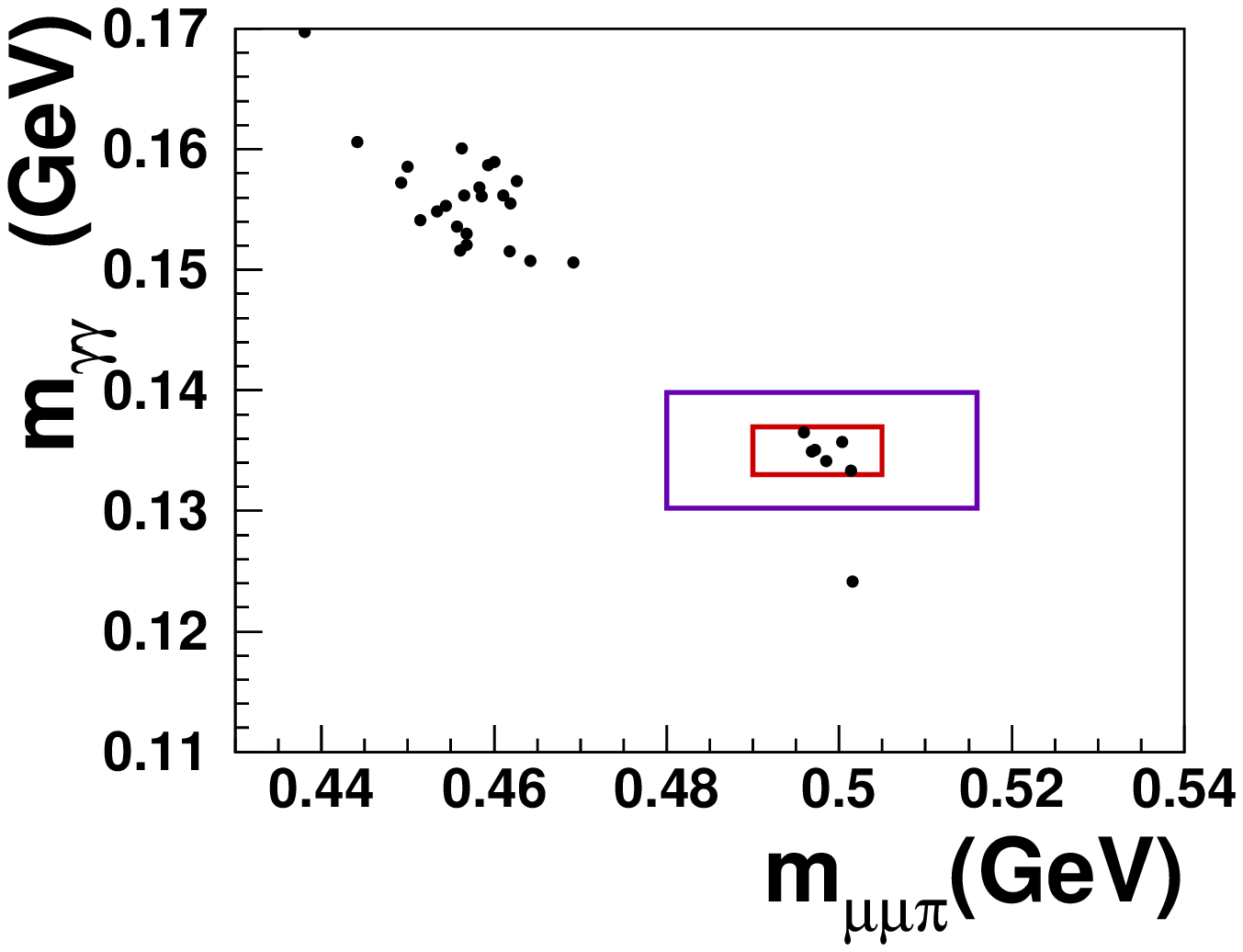}
    \hspace{0.25in}
    \epsfxsize=3.0in
  \epsffile{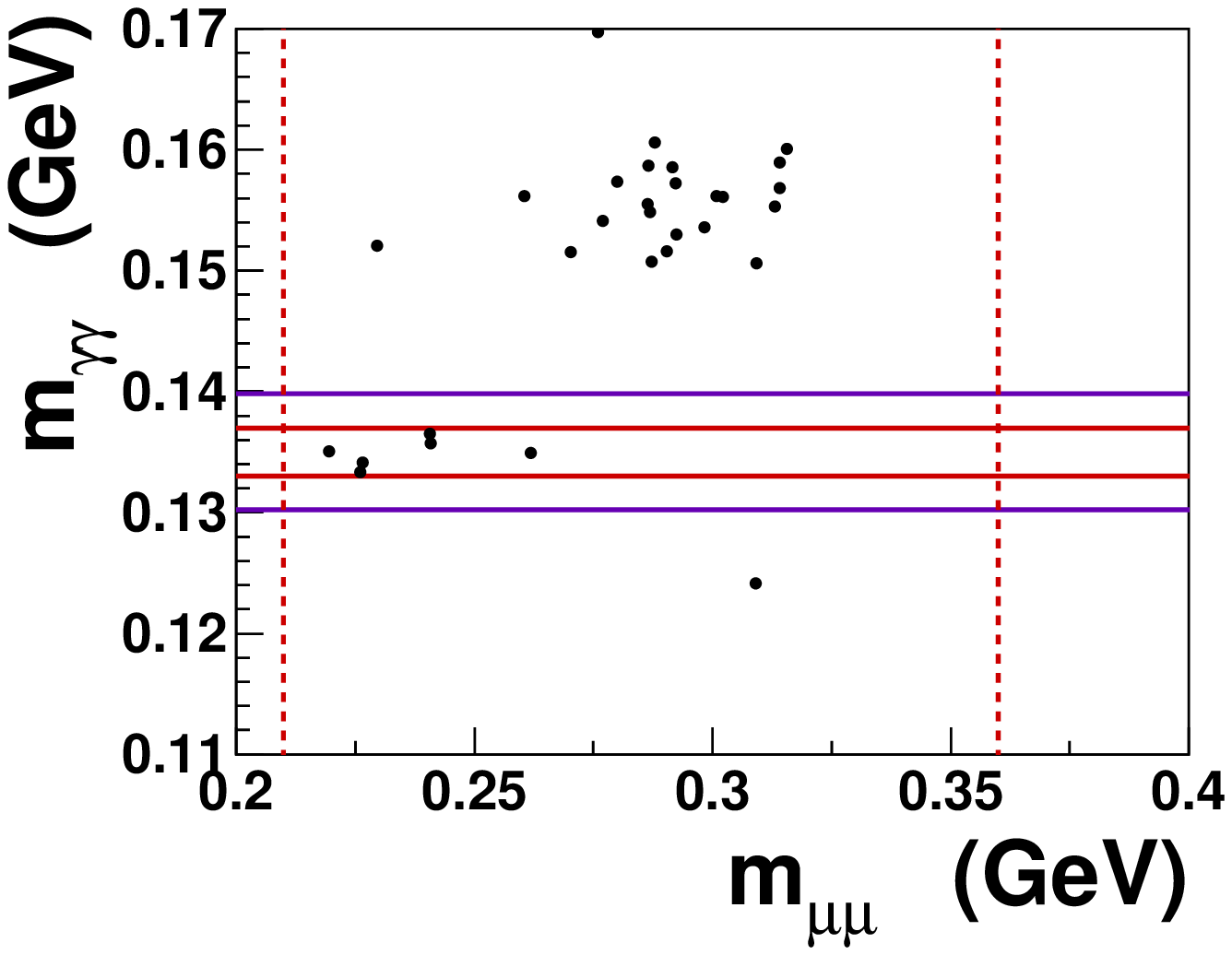}
}
}
\caption{
Scatter plot for the events passing all the cuts described in the text:
(a) for the 
$m_{\gamma \gamma}$
versus 
$m_{\mu\mu\pi}$ plane and
(b) for the 
$m_{\gamma \gamma}$
versus 
$m_{\mu\mu}$ 
plane.
The $2.5 \sigma$ and the $6 \sigma$ signal and control regions
and the $m_{\mu\mu}$ kinematic limits are also shown.
\label{fig:events}}
\end{figure}

Track pairs reconstructed in the spectrometer were classified as muons
if the following conditions were satisfied:
\begin{itemize}
\item 
either no LKr cluster was associated to each track, or the
      energy of any associated cluster was less than 1.5\,GeV.
A cluster was associated with a track if the separation of the cluster and 
the track was less than 6\,cm
and the absolute time difference between the track time 
and the cluster time was less than 4.5\,ns;
\item each track had associated hits in the first two planes ($M1X$ and $M1Y$) of the MUC;
\item the total energy in the hadron calorimeter was less than 10~GeV;
\item for each track, the time measured by either the drift chambers or the  trigger 
hodoscope and the time measured by the muon detector
did not differ by more than 4.5\,ns.
In about 95\% of the events
the track time was given by the trigger hodoscope,
while for the remaining events
the time was given by the drift chambers;
\item the average of the track times and the average time in the
muon detector did not differ by more than 3\,ns;
\item the tracks had opposite charge;
\item the individual track momenta were above 10~GeV in order to have high muon identification efficiency;
\item the transverse momentum with respect to the kaon line of flight
of each track was below 0.180~GeV;
\item the distance between the impact points of the two tracks
at the first drift chamber was greater than 10\,cm
in order to have high trigger vertex reconstruction efficiency;
\item the ratio of the track momenta was larger than 1/3 or smaller
than 3 in order to reduce the background from $\Lambda$ and $\bar{\Lambda}$  decays 
coming either directly from the target or from 
$\Xi^0 (\bar{\Xi^0})
\rightarrow \Lambda (\bar{\Lambda})  \pi^0$ decays;
\item  the reconstructed invariant mass assuming $\Lambda \rightarrow p \pi^{-}$
(or $\bar{\Lambda}$) was   not compatible with the $\Lambda (\bar{\Lambda})$ mass
( 1.1115\,GeV $< m_{p\pi} <$ 1.1200\,GeV);
\item  0.21\,GeV $< m_{\mu\mu} <$ 0.36\,GeV.
\end{itemize}
\noindent
Clusters reconstructed in the LKr calorimeter were classified as photons
if the following conditions were satisfied:
\begin{itemize}
\item the energy of the cluster was greater than 3\,GeV and less than 100\,GeV;
\item the distance to the nearest cluster was greater than 10\,cm in order to minimize the
effect of energy sharing on cluster reconstruction; 
\item the distance to the nearest extrapolated track was greater than 20\,cm.
\end{itemize}
\noindent
In addition, a $\pi^0$  candidate satisfied the following conditions:
\begin{itemize}
\item  the total energy of the clusters was greater than 20~GeV;
\item the transverse momentum with respect to the kaon line of flight
of the $\pi^0$ candidate was between 0.05 and 0.25\,GeV.
\end{itemize}
\noindent
Three quantities related to the decay vertex were computed:
\begin{itemize}
\item {\it total vertex, $z_K$}.

To compute the longitudinal vertex position, $z_K$,
the kaon mass was assumed and the kinematical information from
all the particles involved in the decay was used.
The distance from
the vertex to the LKr calorimeter, $d_{K}$, was calculated as:
\begin{equation}
\label{eq:zk}
\small{
d_{K}=\sqrt{ 
\frac{
\sum_{i\ne j}p_ip_jd_{ij}^2}
{m_{K}^{2} - 2 m_{\mu}^{2}-2\sum_{i\ne j}(E_iE_j-p_ip_j)} 
 } 
}
\end{equation}
where $d_{ij}$ is the separation between the $i$'th and $j$'th particles 
in the LKr plane
after linearly extrapolating charged tracks from before the spectrometer magnet,  
and $E_i$ and $p_i$  are the energy and the momentum of the $i$'th particle.
The vertex is given by
\mbox{$z_K = z~(\mathrm{LKr~ position}) - d_{K}$}.
This formula follows from energy and momentum conservation applied to the
decay $K^0\rightarrow\mu^+\mu^-\gamma\gamma$.
The approximation
$\cos \alpha_{ij} \sim 1 - \frac{1}{2}\,(d_{ij}/d_{K})^2$
was made, where $\alpha_{ij}$ is the angle defined between the
momenta of each  pair of particles.

\item {\it $\pi^0$ vertex, $z_{\pi^0}$ }.

The $\pi^0$ vertex position along the beam direction, $z_{\pi^0}$,
was computed in a similar way to the total vertex,
but used only the two photon clusters and imposed the
$\pi^0$ mass instead of the kaon mass as a constraint.

\item {\it track vertex, $z_{ch}$}.

The track vertex was computed by finding the position of the closest 
distance of approach\,(CDA) between the two tracks.
The CDA was required to be less than 1.5\,cm in order to reject muons from pion decay.
\end{itemize}
\noindent
Further cuts were applied:
\begin{itemize}
\item on the assumption that  the observed event was   a kaon decay, its 
proper lifetime was calculated using 
the $z_K$, $z_{\pi^0}$ and $z_{ch}$ vertices,
taking the end of the final collimator as the origin.
All three calculated proper lifetimes
were required to be between 0 and 3 $K_S$ mean lifetimes.

\item the energy-weighted centre-of-gravity\,(COG) of the event
was required to be less than 5\,cm
in order to further reduce the background from accidental activity. 
The COG was defined as  
COG=$\sqrt{(\sum_i E_i x_i)^2+(\sum_i E_i y_i)^2}/\sum_i E_i$, 
where $x_{i}$ and $y_{i}$ 
are the coordinates of the  $i$'th particle 
in the LKr plane after linearly extrapolating charged tracks 
from before the spectrometer magnet,  
and $E_i$  is  the energy of  the $i$'th particle.

\item events with extra tracks 
within the trigger window  or extra clusters 
in the readout window were rejected.

\item 
the time difference $\Delta t$ between the average track time
and the average time of the 
LKr clusters forming the $\pi^0$ was required to be less than 1.5\,ns.

\item fiducial cuts: tracks were required to be at least 12\,cm from the centre 
of the drift chambers
and clusters in the LKr were required to be at least 15\,cm from the centre of the LKr,
11\,cm from its outside borders, and 2\,cm from any defective cell.
\end{itemize}
\noindent
After applying all these cuts, the signal and control regions were unmasked.
Six events were found in the signal region, and none in the control region
(see Fig.\,\ref{fig:events}).
The kinematical quantities for these events are summarised  in
Table~\ref{tab:events}.

\subsection{Background}
\label{sec:back}
There are two categories of background:\,
{\it physical background} and {\it accidental background}.
Physical background was defined as that
due to a single kaon decay, for example
$ K_L \rightarrow \pi^+ \pi^- \pi^0$,
where the two charged pions decayed into muons.
Accidental background was caused by the overlap of particles coming
from two separate decays that happened to be in time and faked the signal,
for example, an overlap between the decays  $K_L \rightarrow \pi \mu \nu$, 
where $\pi \rightarrow \mu \nu$,
and $ K_S \rightarrow \pi^0 \pi^0$,
where two photons missed the detector.
The accidental background contained two components:
(a) overlapping fragments from different proton interactions
in the target; and
(b) fragments from associated production ($KK$ or $\Lambda K$) 
due to a single proton interaction. 
The first component could be  estimated through a study of out-of-time 
sidebands.

\subsubsection{Physical backgrounds}
Only two sources of physical background were found to give a significant contribution:

(1) $K_{L,S} \rightarrow \mu^+ \mu^- \gamma \gamma$ decays:\\
The $\KL\rightarrow\mu^+ \mu^- \gamma \gamma$
branching ratio has been measured by the KTeV Collaboration
to be $(1.0^{+0.8}_{-0.6})\times 10^{-8}$~\cite{bib:ktev_mumugg}.
Using Monte Carlo simulation,
the acceptance for $K_{L} \rightarrow \mu^+ \mu^- \gamma \gamma$ decays
in NA48 was found to be $5\times 10^{-3}$.
This small value originates from a geometrical acceptance of about 20\%
together with the low  probability that
the invariant  mass of the  two photons is consistent with the $\pi^0$ mass.  
The $K_S$ decay into the same final state was taken into account by 
assuming equal decay rates for $K_S$ and $K_L$.
The total  background was estimated to be 
$0.04^{+0.04}_{-0.03}$ event. 

\begin{figure}[hbtp]
  \vspace{9pt}
  \centerline{\hbox{ \hspace{0.0in}
    \epsfxsize=3.0in
    \epsffile{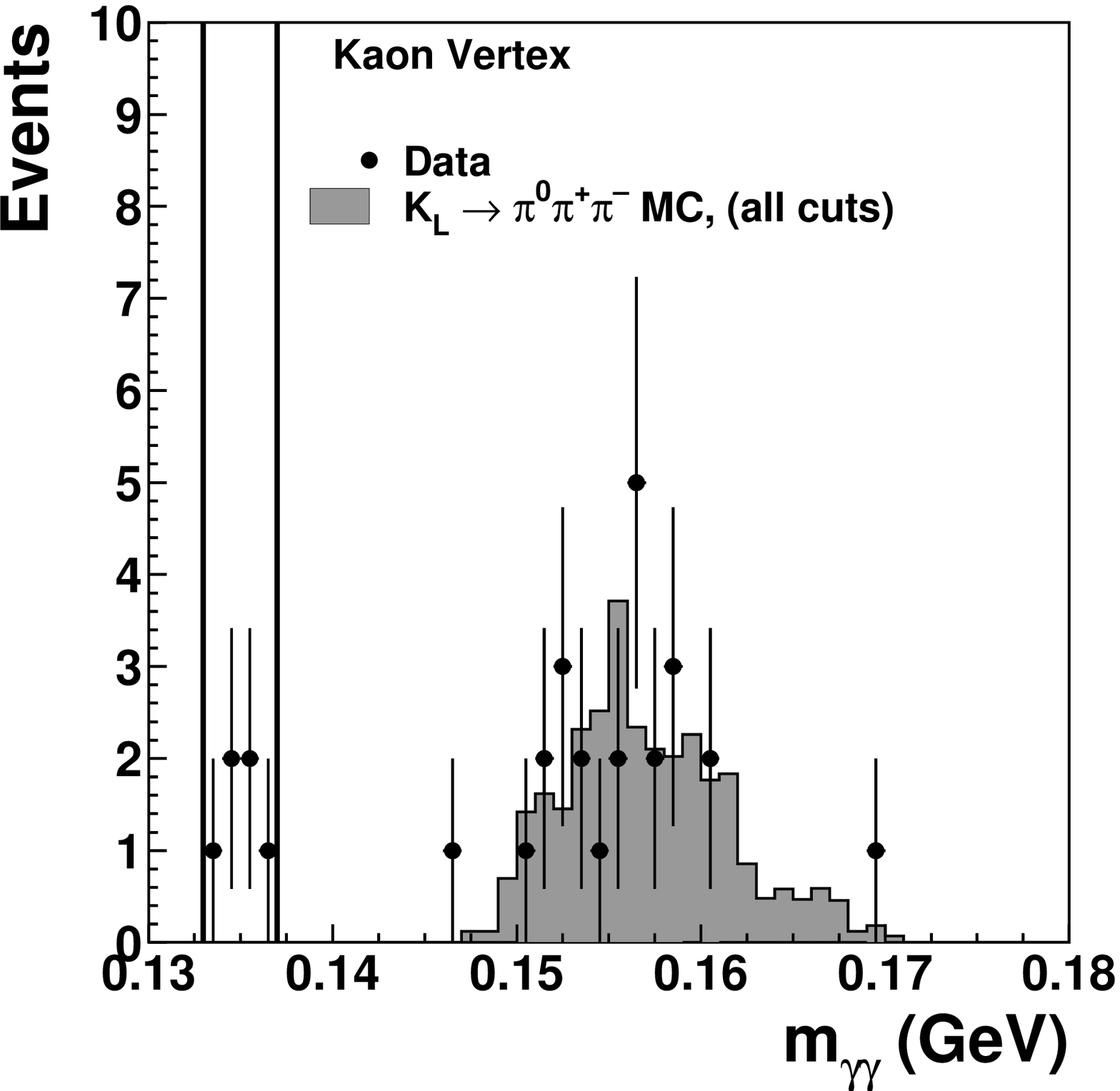}
    \hspace{0.25in}
    \epsfxsize=3.0in
    \epsffile{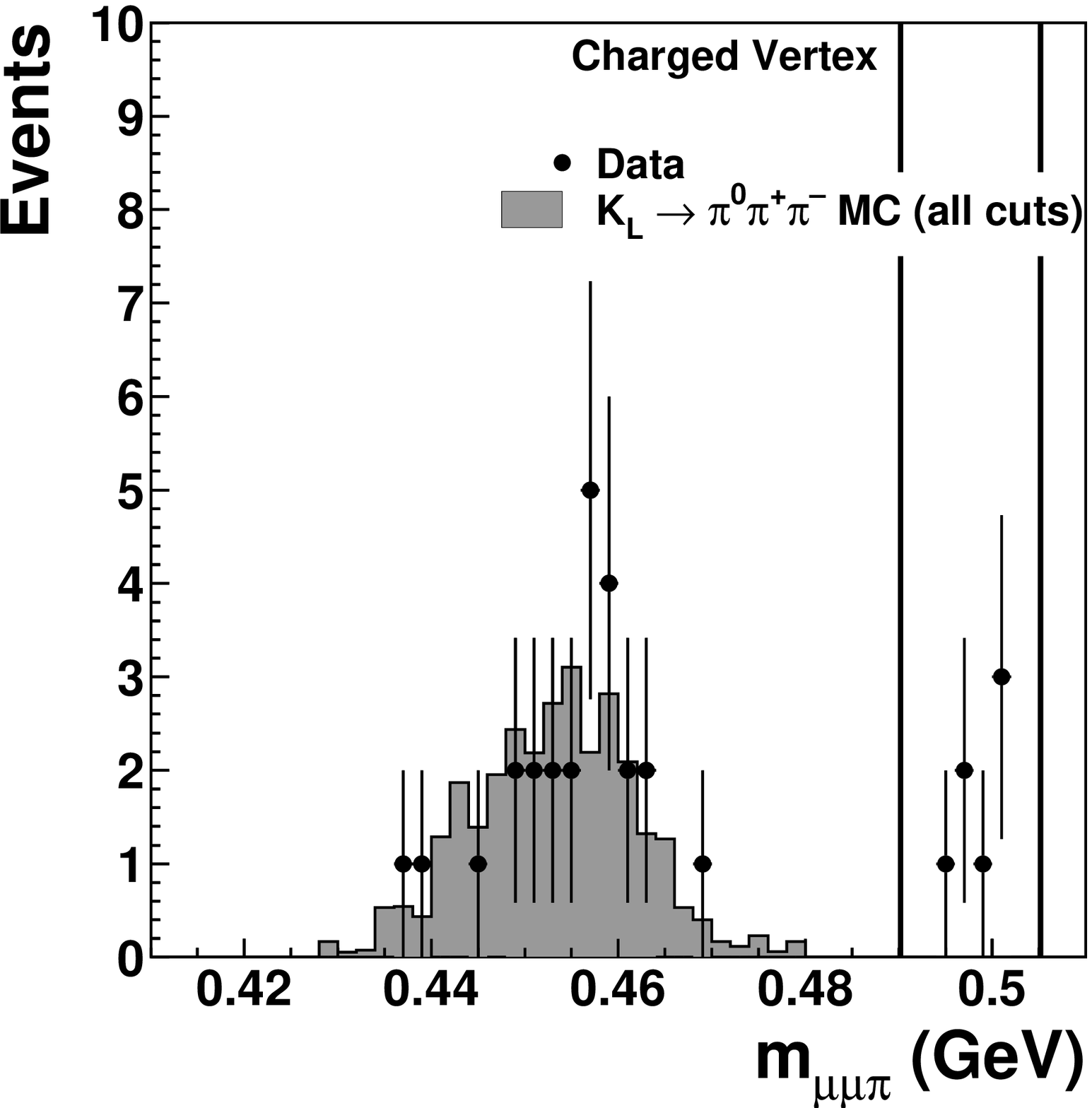}
}
}
  \centerline{\hbox{ \hspace{0.0in}
    \epsfxsize=3.0in
    \epsffile{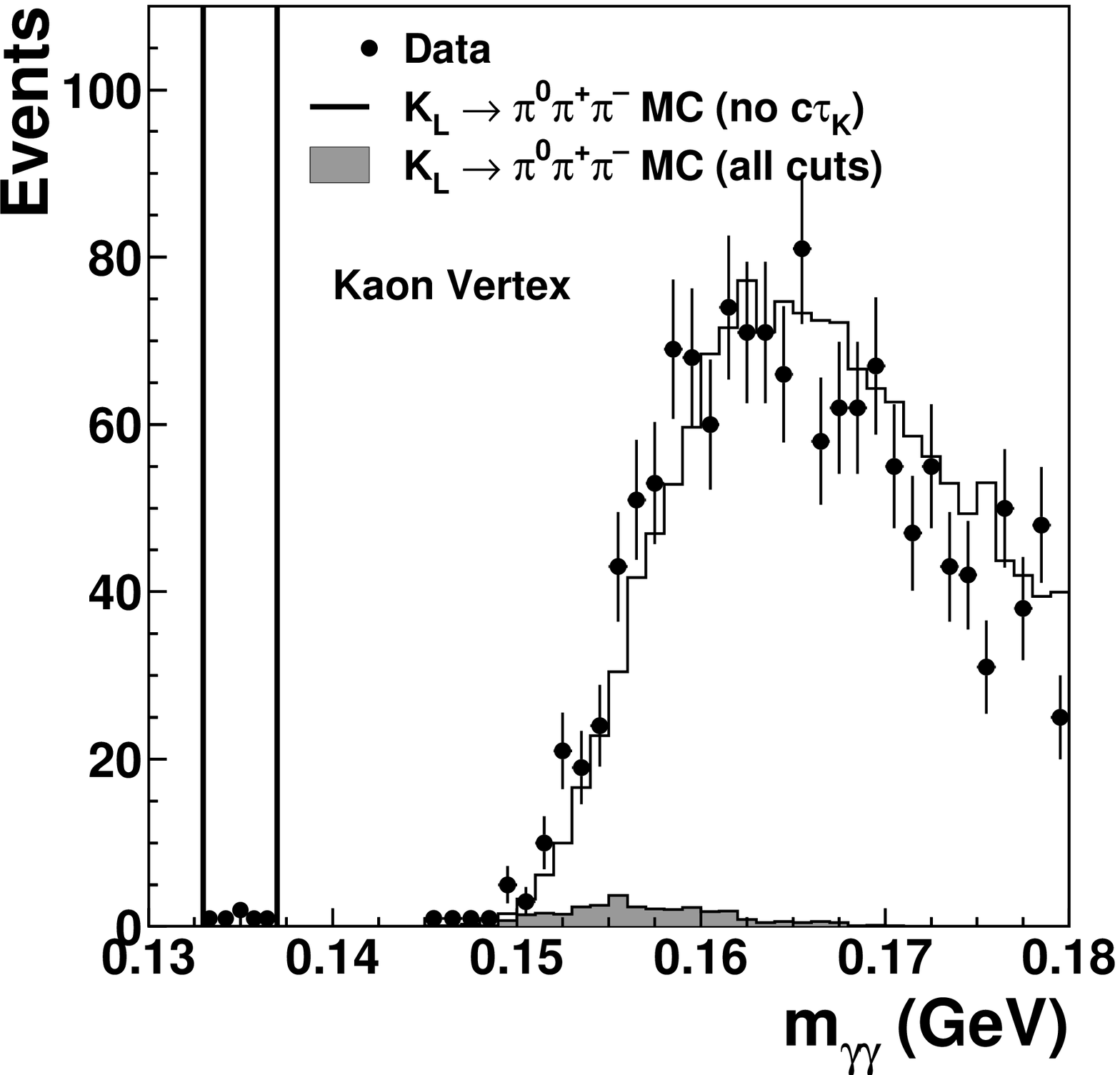}
    \hspace{0.25in}
    \epsfxsize=3.0in
    \epsffile{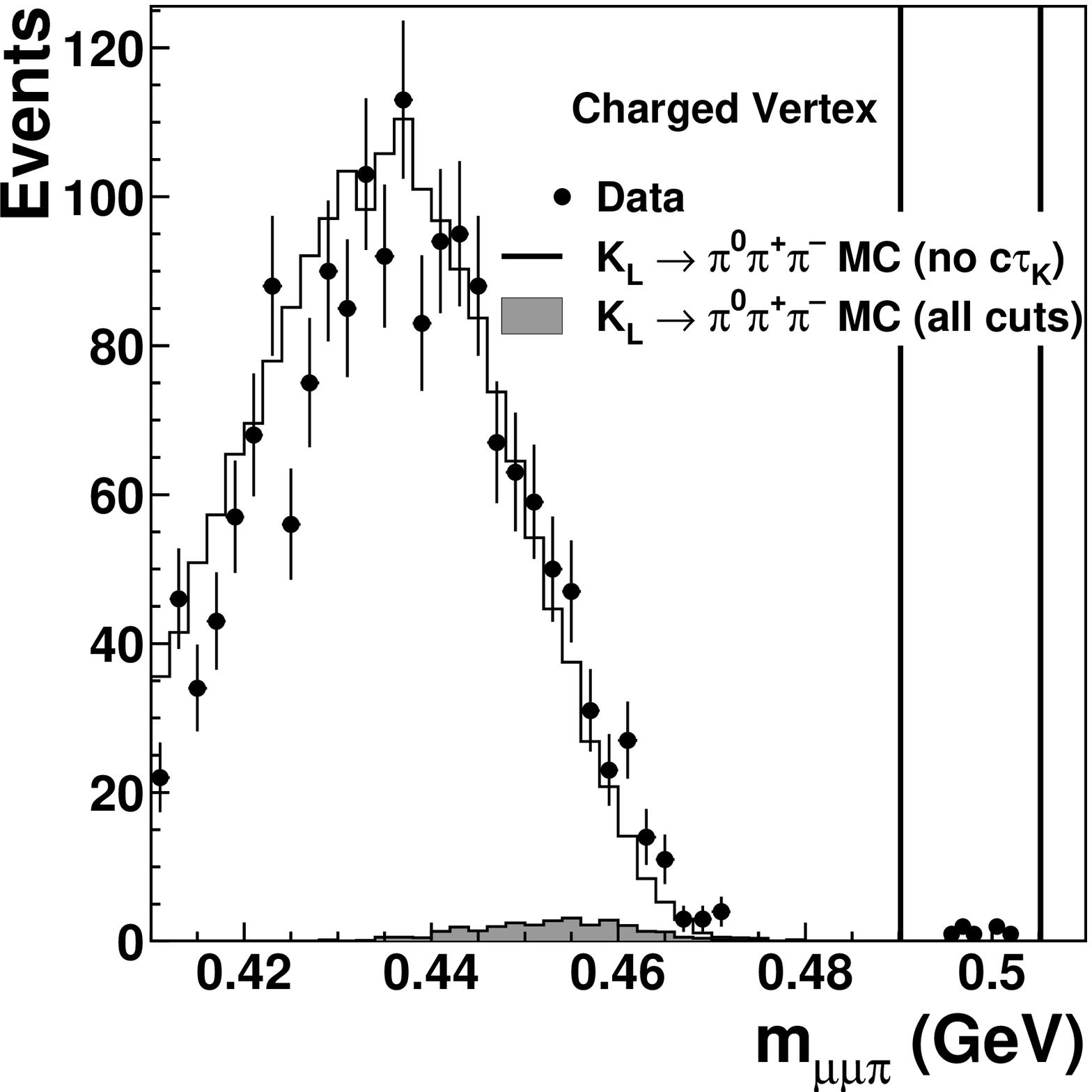}
}
}
\caption{
Distributions of $\mgg$
and  $m_{\mu\mu\pi}$, with the cuts on  $\mgg$ and  $m_{\mu\mu\pi}$ removed,
for data (points with error bars)
and $K_L\rightarrow \pi^+\pi^-\pi^0$ Monte Carlo (histograms).
The  lower plots do not 
include the $K_S$ proper lifetime cut calculated from $z_K$.
The vertical lines indicate the 2.5\,$\sigma$ signal region.
\label{fig:data_mc}}
\end{figure}

(2) $K_{L} \rightarrow \pi^+ \pi^- \pi^0$ decay, where both pions have
decayed in flight: \\
For this background, the reconstructed total mass $m_{\mu\mu\pi}$
    lies below the known kaon mass as a consequence of the missing energy
    of the neutrinos and the use of the muon mass rather than the pion mass
    in the reconstruction. For the same reasons, the $z$ vertex 
position\,(Eq.~\ref{eq:zk})
computed by imposing the $K^0$ mass constraint is pushed upstream
    from the actual vertex position of the $\pi^+\pi^-\pi^0$ decay.
    The shift in the $z$ vertex position leads to a reconstructed
    two-photon mass $m_{\gamma\gamma}$ lying above the known $\pi^0$ mass.
    These effects are clearly seen in 
Fig.~\ref{fig:data_mc},
where the $m_{\gamma\gamma}$
    and $m_{\mu\mu\pi}$ distributions are compared for data and Monte Carlo,
    after applying all cuts and
    after removing the $K_S$ proper lifetime cuts calculated from $z_K$.
    The Monte Carlo gives a good description of the data.

An extended control region was defined
in the ($m_{\mu\mu\pi},m_{\gamma\gamma}$) plane,
between $6\sigma$ and $12\sigma$ around
the $\pi^0$ and $K_S$ masses. No data event was found in this region.
A Monte Carlo sample equivalent to 24 times the 2002 data
was used to estimate the background.
The cuts on the $K_S$ lifetime, on the transverse momenta of the muons
and the pion and on the total energy of the two photons
were relaxed, giving 11 Monte Carlo 
events in the extended control region and no event in the signal region.
This led to a background estimate of $<2.44/24 = 0.10$ at 90\%CL.
It was checked that the transverse momentum and photon
    energy cuts did not significantly affect the shape of the relevant
    background distributions. 
After re-applying all cuts except the $K_S$ lifetime cut,
2 of the 11  Monte Carlo events remained in the extended control 
region. This led to an additional extrapolation factor of $2/11 = 0.18$, 
giving an overall background estimate for this mode of $<0.018$ event
 at 90\%CL.

\subsubsection{Accidental backgrounds}
%

\begin{figure}[hbtp]
  \vspace{9pt}
  \centerline{\hbox{ \hspace{0.0in}
    \epsfxsize=3.0in
   \epsffile{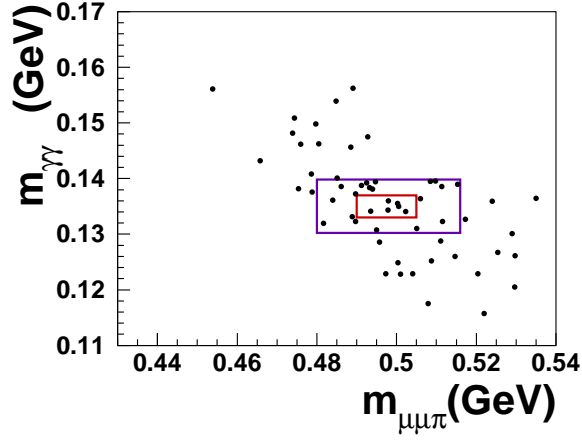}        
  }
  }
\caption{Scatter plot of
$m_{\gamma \gamma}$
versus 
$m_{\mu\mu\pi}$
for events in the out-of-time window.
The $2.5 \sigma$ and the $6 \sigma$ signal and control regions are also
shown.
\label{fig:out_of_time}}
\end{figure}

Accidental background was studied using data with relaxed timing requirements.
As a consequence of
the trigger system, an asymmetric timing window was used.
Events in the time sidebands ($-115$\,ns $\le \Delta t \le -3$\,ns) and 
(3\,ns $ \le \Delta t \le $ 60\,ns) were used to extrapolate the background 
from the out-of-time to the in-time 
($-1.5$\,ns $\le \Delta t \le +1.5$\,ns) signal region. 
$ \Delta t$ was defined as the time difference between the average track 
time and the average time of the LKr clusters forming the $\pi^0$.
Given the number of $K_S$ decays in the fiducial volume
and the spill length, the probability that two events overlap in 
time was $\sim 1\times 10^{-4}$.
The out-of-time events are shown in Fig.~\ref{fig:out_of_time}. 
The readout system introduced a non-uniformity in the
event time distribution which changed the effective width of the 
out-of-time window from ($175-6$) to ($125-6$) ns.
Due to additional triggering effects,
events in this window were recorded only
for the $2\mu_{\mathrm{tight}}$ component of the trigger.
An extrapolation factor of 1.27 was applied to take into account 
the relative acceptances of the $2\mu_{\mathrm{tight}}$ and $2\mu_{\mathrm{loose}}$ triggers (see Sections~\ref{sec:trig_effy} and \ref{sec:acceptance}).
Six events were observed
in the out-of-time signal region, five of them in the later data-taking 
period when the $2\mu_{\mathrm{loose}}$ trigger was active, leading to 
an overall
accidental background estimate of $((5\times 1.27)+1)\times 3/(125-6) = 0.18$ 
event.

Studies have shown that the majority of 
the accidental background was due to $\kspipic$ or $\klpimunu$ decays
which were in time with two photons from a $\kpiopio$ decay.
The background for \mbox{$m_{\mu\mu} <$ 0.3\,GeV} was dominated 
by $\klpimunu$ decays,
while for \mbox{0.30\,GeV $< m_{\mu\mu} <$ 0.36\,GeV}, the background was
dominated by\break $\kspipic$ decays.
Background where the two charged tracks came 
from different decays was examined by looking for events where the two
tracks had the same charge.  No same-sign event was found in the 
signal region,
consistent with the background estimate above.
The accidental background from associated production was found 
to be negligible.

\subsubsection{Background summary}
The total background was estimated to be  
$0.22^{+0.18}_{-0.11}$  event in the signal region.
The significant background contributions are summarised in Table~\ref{tab:back}.

\subsection{Trigger efficiency}
\label{sec:trig_effy}

{\it Hardware trigger:}\\
The trigger efficiency was dominated by the geometrical acceptance of the
two-muon signal and by the charged vertex requirement.
The efficiencies of the other trigger components were measured,
but no correction was necessary since   
they  were also present in the normalisation trigger.

${K_L} \rightarrow \pi^+\pi^- \pi^0$ events were used to measure the efficiency
of the hardware coordinate builders and the microprocessors that reconstructed
the tracks and determined the vertex of the decay. The efficiency of 
this trigger component was  found to be $0.910\pm 0.001$.

\begin{figure}[hbtp]
  \vspace{9pt}
  \centerline{\hbox{ \hspace{0.0in}
    \epsfxsize=3.0in
  \epsffile{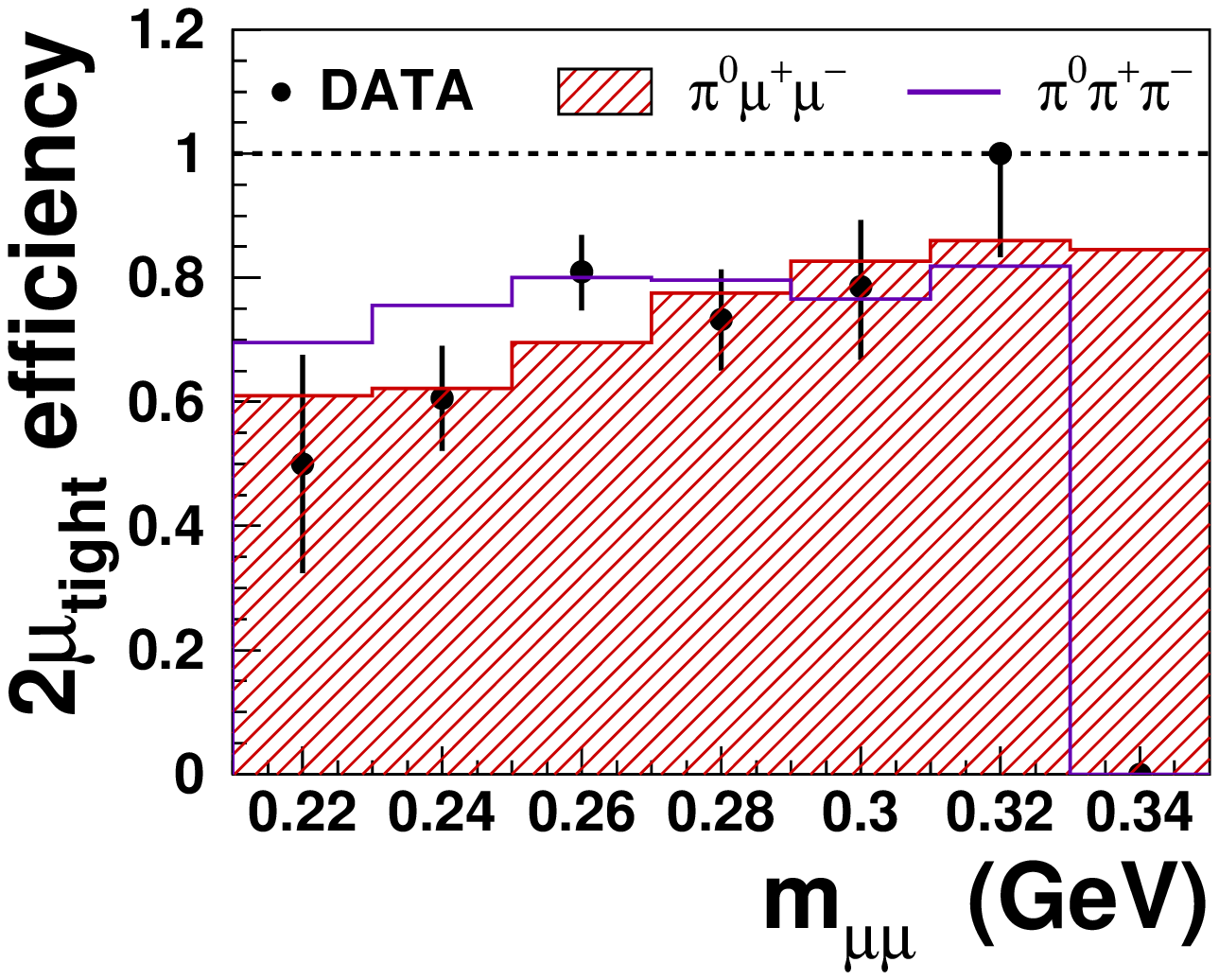} 
    \hspace{0.25in}
    \epsfxsize=3.0in
  \epsffile{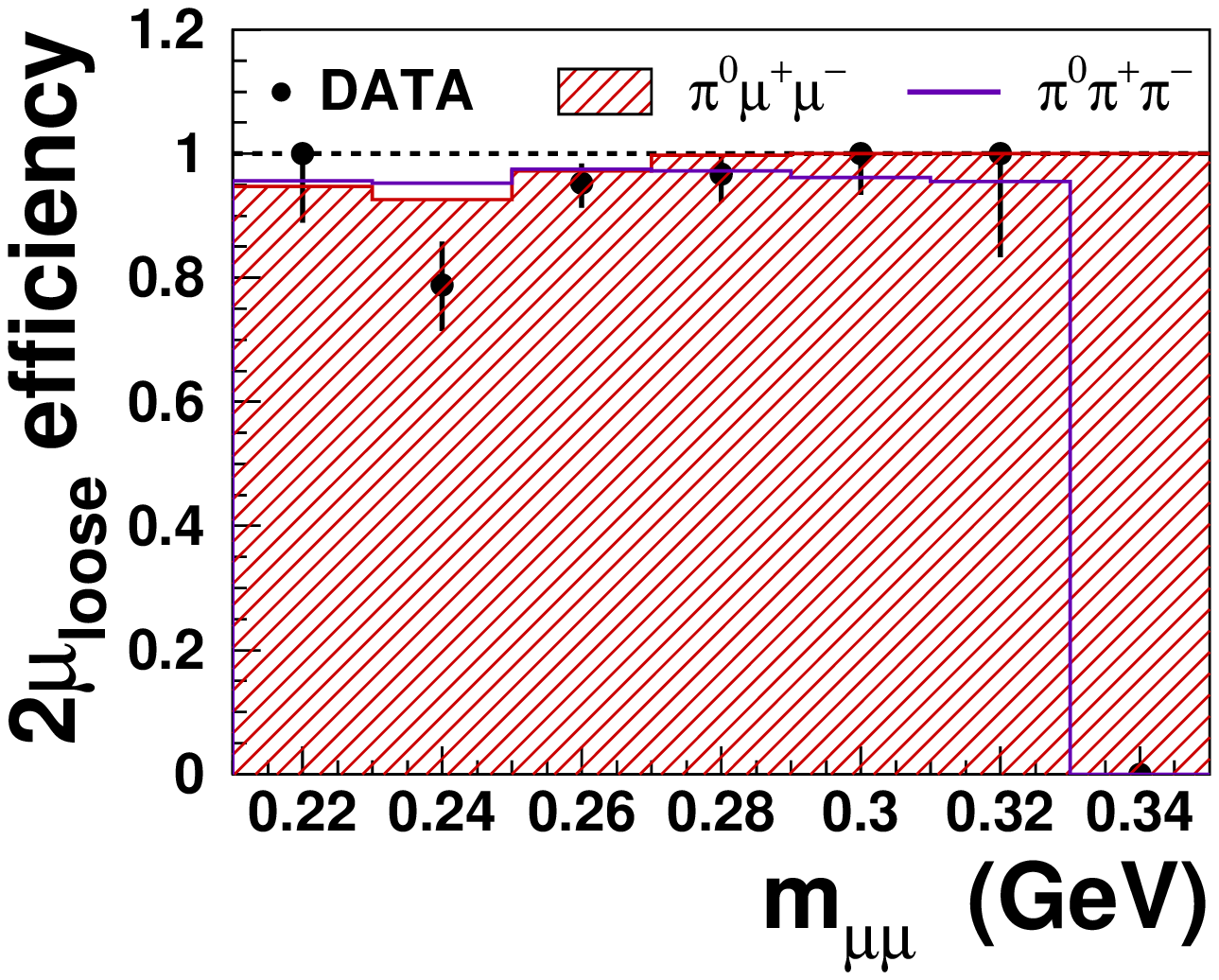} 
    }
  }
\caption{Trigger efficiency measured from data (points with error bars)
and estimated from Monte Carlo (histograms) for the two-muon signal 
component of the trigger, as a function of $m_{\mu\mu}$: 
(a) for $2\mu_{\mathrm{tight}}$, 
and (b) for
$2\mu_{\mathrm{loose}}$. 
\label{fig:trigger}
}
\end{figure}

The efficiency of the two-muon signal component of the trigger
was estimated from ${K_L} \rightarrow \pi^+\pi^- \pi^0$ decays
where both charged pions decayed to muons.
The kinematics of these events
is very similar to that  of
${K_S} \rightarrow \pi^0\mu^+\mu^- $.
The efficiencies of the $2\mu_{\mathrm{tight}}$ and $2\mu_{\mathrm{loose}}$ 
signals are  shown in Fig.~\ref{fig:trigger} as a function of 
$m_{\mu\mu}$.
A large track separation was effectively imposed by the $2\mu_{\mathrm{tight}}$ 
trigger requirement,
resulting in a lower trigger efficiency at low $m_{\mu\mu}$.
As shown in Fig.~\ref{fig:trigger},
the dependence of the trigger efficiency on $m_{\mu\mu}$
was consistent with that expected from Monte Carlo simulation.

The $2\mu_{\mathrm{tight}}$ and $2\mu_{\mathrm{loose}}$
trigger efficiencies for $\kspimumu$ decays
were found to be $0.73\pm 0.03$ and $0.93\pm 0.02$, respectively.
Taking into account the relative amount of data recorded with the
$2\mu_{\mathrm{tight}}$ and  $2\mu_{\mathrm{loose}}$ triggers,
and including the vertex-finder trigger efficiency, 
the overall hardware trigger efficiency correction
 was determined to be $0.77\pm 0.02$.

{\it Software trigger:} \\
A small fraction of unbiased events was flagged by the software trigger, 
but not rejected. Using these events, 
the efficiency for the software trigger described in
Section~\ref{sec:t_and_r} was determined to be $1.00^{+0.00}_{-0.02}$.

\subsection{Signal Acceptance}
\label{sec:acceptance}

The signal acceptance was estimated using Monte Carlo simulation.
The amplitude for the decay 
was taken to be a vector matrix element
of the form~\cite{D'Ambrosio:1998yj}: 
\begin{equation}
A[K_S \rightarrow \pi^0 \mu^+  \mu^- ] \propto
 W(z) (p+p_\pi)^{\mu}
{\bar u}_l (p_-) \gamma_{\mu} v_l (p_+),
\label{eq:matrix}
\end{equation}
\par\noindent
where $p$, $p_\pi$, $p_-$ and $p_+$ are the four-momenta
of the kaon, pion, muon and anti-muon, respectively, $z=m_{\mu\mu}^2/m_K^2$, 
and $W(z)$ is a form factor.
As explained in~\cite{D'Ambrosio:1998yj}, the $W(z)$ dependence on $z$
vanishes to lowest order and for this analysis was represented by
the first order polynomial $W(z) =  a_S + b_S z$.

The response of the LKr and HCAL detectors to muons
is not well simulated by the Monte Carlo
and the effect of the LKr and HCAL energy cuts  on the muon identification 
efficiency
was therefore  studied using data. From a special run with a muon beam,
the mean $2\mu$ identification efficiency was determined to be $0.96\pm 0.02$.
The dependence of the trigger efficiency on $m_{\mu\mu}$
was also determined from data,
as described in the previous section.

\begin{figure}[hbtp]
  \vspace{9pt}
  \centerline{\hbox{ \hspace{0.0in}
    \epsfxsize=3.0in
  \epsffile{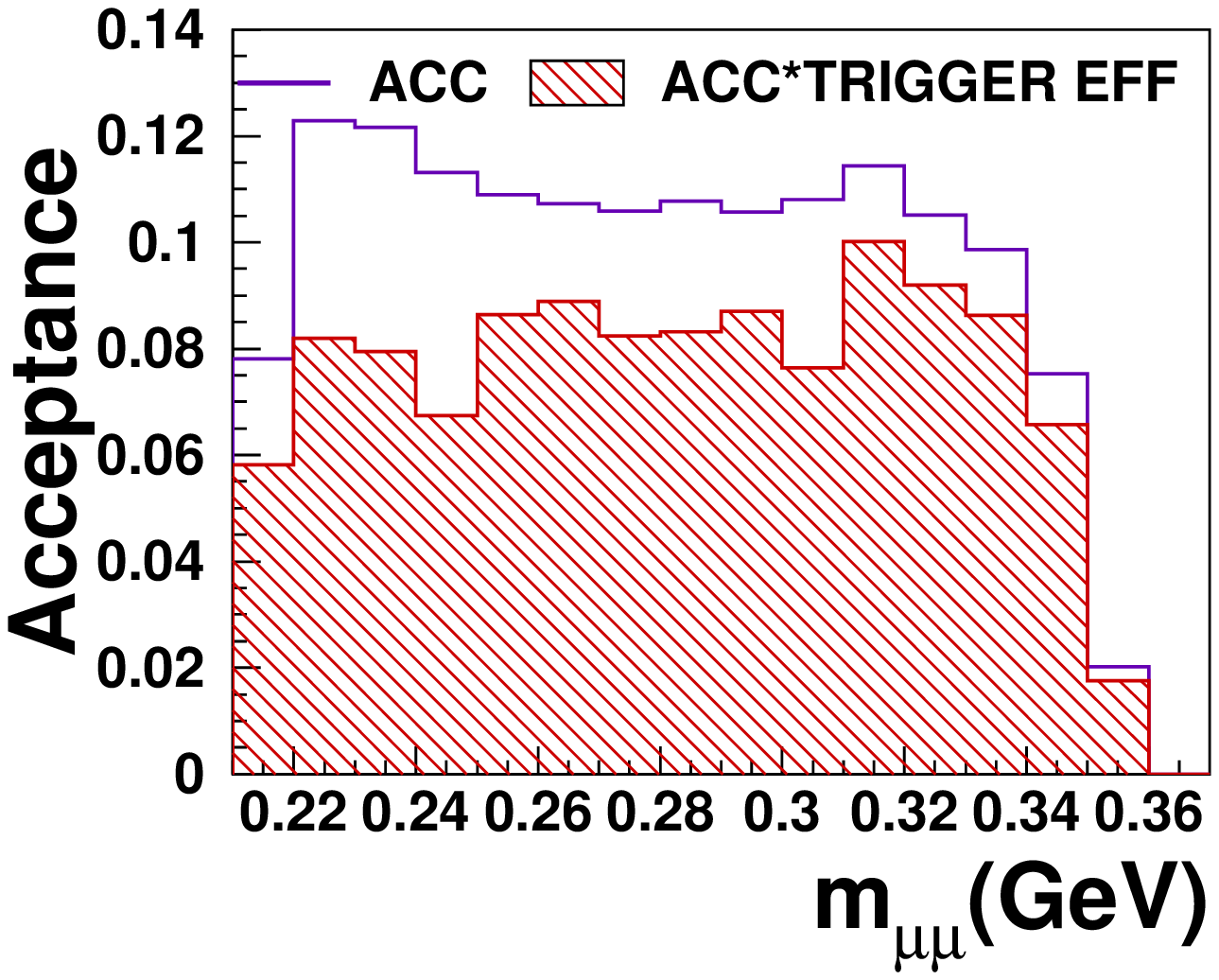}
    \hspace{0.25in}
    \epsfxsize=3.0in
  \epsffile{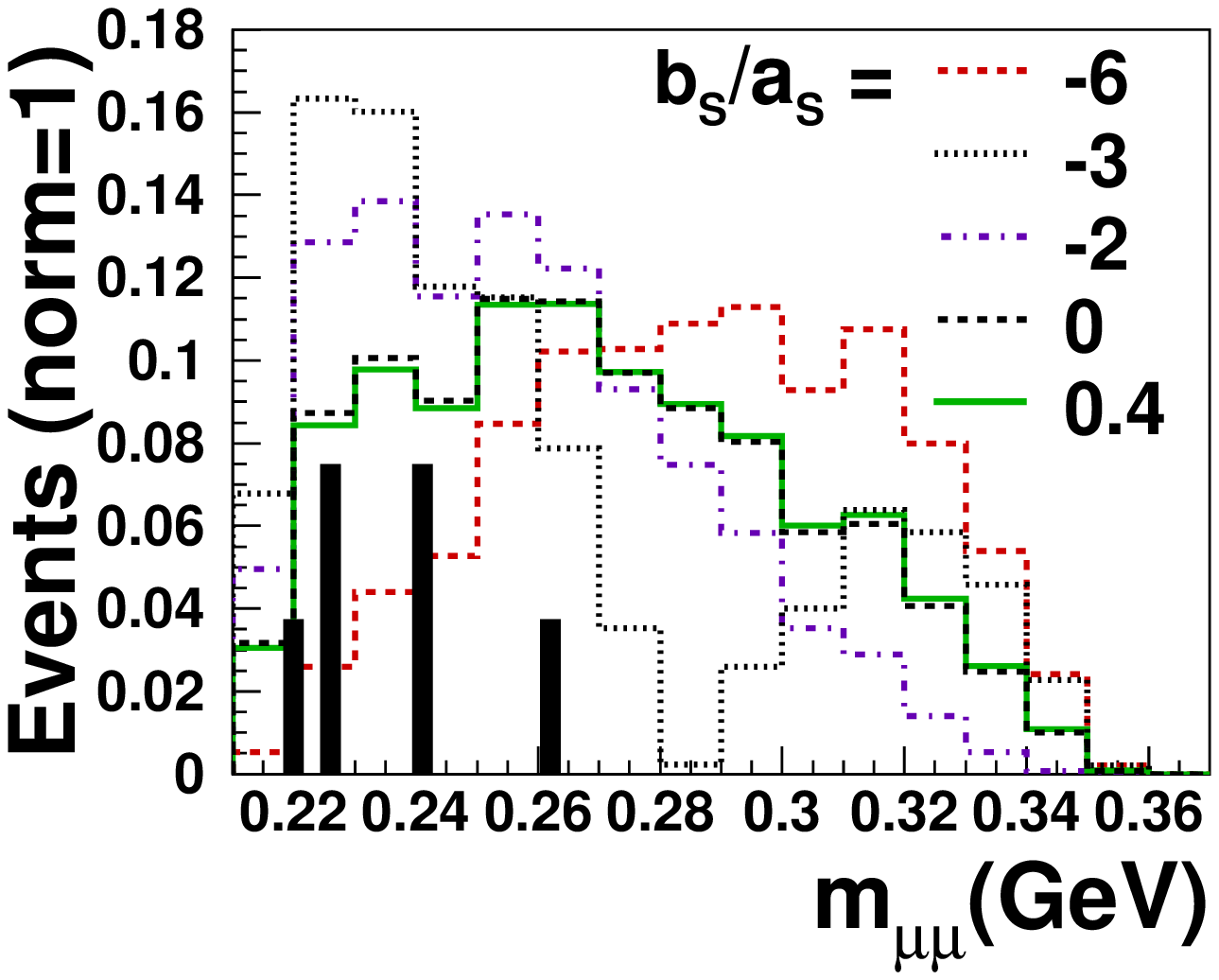}
    }
  }
\caption{(a) Acceptance as a function of $m_{\mu\mu}$,
including (hatched histogram) and not including (open histogram)
the trigger efficiency; 
(b) Expected $m_{\mu\mu}$ distributions of accepted events for
several values of $b_S/a_S$
after taking into account the trigger efficiency.
The $m_{\mu\mu}$ distribution of the six signal events
is also shown.
\label{fig:accep}
}
\end{figure}

The acceptance as a function of $m_{\mu\mu}$
is shown in Fig.~\ref{fig:accep}(a),
before and after applying the muon identification and trigger efficiency corrections (see Table~\ref{tab:acceptance_m}).
The reduced      acceptance at low $m_{\mu\mu}$ after applying these corrections
is due to the $2\mu_{\mathrm{tight}}$ component of the trigger.

The expected $m_{\mu\mu}$ distribution,
taking into account the muon identification and trigger efficiency corrections,
is shown in Fig.~\ref{fig:accep}(b)
for several values of $b_S/a_S$,
including a unit form factor, $b_S/a_S=0$,
and $b_S/a_S=0.4$ as suggested by the Vector Meson Dominance\,(VMD) model.
The reconstructed dilepton mass distribution of  the six signal events is also
shown in Fig.~\ref{fig:accep}(b)
and is consistent with expectation.

\begin{figure}[hbtp]
  \vspace{9pt}
  \centerline{\hbox{ \hspace{0.0in}
    \epsfxsize=4.0in
    \epsffile{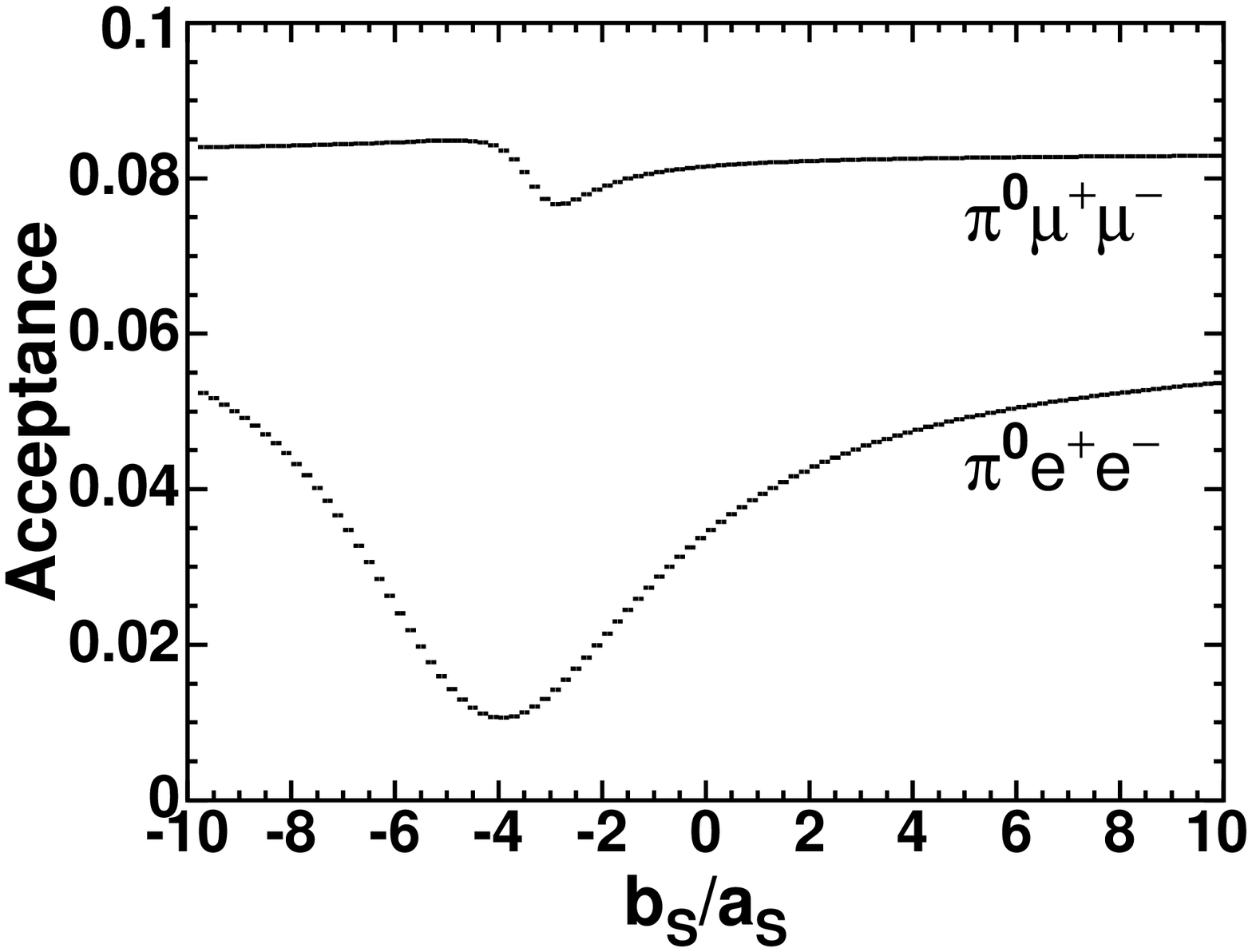}
    }
  }
\caption{
Overall acceptance as a function of $b_S/a_S$
for the $\kspimumu$ (upper curve) and $\kspiee$ (lower curve) channels.
\label{fig:accep2}
}
\end{figure}

The dependence of the overall acceptance on the value of $b_S/a_S$
is shown in Fig.~\ref{fig:accep2}.
The variation in acceptance seen in the region
$-5\lesssim b_S/a_S \lesssim -2$
arises because the form factor develops a minimum within the kinematically allowed
range of $m_{\mu\mu}$ (see Fig.~\ref{fig:accep}(b)).
The maximum variation in the acceptance over a wide range of
$b_S/a_S$ was used to determine the systematic uncertainty.


Taking into account the dependence of the trigger efficiency
on $m_{\mu\mu}$ and the muon identification efficiency,
the overall acceptance was found to be
\begin{eqnarray}
0.081\pm 0.002 (stat) \pm 0.004(syst)
\end{eqnarray}
for kaons in the energy range 60 to 200\,GeV decaying in the first 
three $K_S$ lifetimes after the collimator,
where a unit form factor has been assumed.
For comparison, the acceptance before applying the trigger and muon identification
corrections was 0.109\,.

\subsection{Normalisation.}

The \KS\ flux was estimated using 132 million  $K_S\rightarrow \pi^+\pi^-$ 
events from a minimum bias trigger.  Only events that had one
vertex, two tracks and no additional clusters outside a modified 
in-time window were used.
Events where the charged pions had decayed were not rejected,
and an $E/p$ cut of less than 0.95 was used.

The flux was corrected for $\pi^+\pi^-$ acceptance\,($0.4579\pm0.0004$), 
software trigger efficiency ($0.9989\pm 0.0004$),
and the $E/p$ cut ($0.9932\pm 0.0001$),
where the errors are statistical only.
No correction was needed for the hardware trigger efficiency, 
for inefficiencies in the chambers and track reconstruction,
or for losses due to multi-vertex and accidental activity since 
these also applied to the signal channel.
To estimate the systematic uncertainty, 
the cuts used to reject accidental activity and 
extra clusters
were varied.
A systematic uncertainty was assigned to take into account the 
3\% variation observed.

After taking into account the prescaling factor of the trigger, 
the total $K_S$ flux  was found to be
 $(2.50\pm 0.08)\times 10^{10}$
in the same fiducial volume as used for the signal channel.

\section{Result}
As shown in  Fig.\,\ref{fig:events}, six events were found in the 
signal region, with a background estimate of $0.22^{+0.18}_{-0.11}$ event. 
This is the first observation of the $\kspimumu$ decay.

The kinematic properties of the six $\kspimumu$ candidates
were consistent with those expected based on Monte Carlo simulation of the signal.
It was also checked that the number of events observed when
the principal analysis cuts were relaxed
was consistent with the expected background
and that there was no accumulation of events close to any of these cuts.

Using the information summarised in Table~\ref{tab:br},
the $\kspimumu$ branching ratio was measured to be:
\begin{equation}
\label{eq:br_result}
B(\kspimumu)=[2.9^{+1.5}_{-1.2}({\it stat})\pm {0.2}({\it syst})]\times 10^{-9},
\end{equation}
where the statistical error was obtained using the 
method in Ref.~\cite{Feldman:1998} and
the systematic uncertainty
by combining the individual errors in quadrature.

This result is consistent within error with recent predictions based on 
Chiral Perturbation Theory~\cite{D'Ambrosio:1998yj,Raphael_04}.

\section{Discussion}
\label{sec:Discussion}

\subsection{Test of Chiral Perturbation Theory}

Chiral Perturbation Theory\,(ChPT) can be used to predict the branching ratio for 
$K_S \rightarrow \pi^0 \ell^+ \ell^-$ and
the corresponding  dilepton mass spectrum, $m_{\ell\ell}$.
The measurement presented here tests these predictions and constrains the 
parameters of the model.

The $K_S \rightarrow \pi^0 \ell^+ \ell^-$
branching ratios can be expressed as a function of two parameters, $a_S$ and  
$b_S$~\cite{D'Ambrosio:1998yj}:
\begin{align}
\label{eq:as_ee}
\!\!\!B(K_S \rightarrow \pi^0 e^+ e^-) &=
\left[ 0.01 - 0.76 a_S - 0.21 b_S + 46.5 a_S^2 + 12.9 a_S b_S + 1.44
b_S^2 \right]\times 10^{-10} \\
\label{eq:as_mumu}
\!\!\!B(K_S \rightarrow \pi^0 \mu^+ \mu^-) &=
\left[ 0.07 - 4.52 a_S - 1.50 b_S + 98.7 a_S^2 + 57.7 a_S b_S + 8.95
b_S^2 \right]\times 10^{-11} 
\end{align}
\noindent 
\noindent where  the  total form factor  is 
$W_S(z)=G_F m_K^2 (a_S + b_S z) + W_S^{\pi\pi}(z)$, 
$z=m_{\ell\ell}^2/m_K^2$, $m_K$ is the kaon mass,
$m_{\ell\ell}$ is the invariant mass of the two leptons,  
and $W_S^{\pi\pi}(z)$ is expected to be small.
Assuming VMD,
which predicts $b_S=0.4\,a_S$ \cite{D'Ambrosio:1998yj},
the value of the parameter $|a_S|$ can be obtained from the measurement
of the individual
$K_S \rightarrow \pi^0 \ell^+\ell^-$
branching ratios via the relations~\cite{Buchalla:2003sj}
\begin{eqnarray}
\label{eq:as_vmd_ee}
B(K_S \rightarrow \pi^0 e^+ e^-) &\simeq& 5.2  \times 10^{-9}\, a_S^2,\\
\label{eq:as_vmd_mumu}
B(K_S \rightarrow \pi^0 \mu^+ \mu^-) &\simeq& 1.2 \times 10^{-9}\, a_S^2.
\end{eqnarray}
The VMD model gives an estimate of the ratio of the muon to electron branching ratios
of $0.23$~\cite{D'Ambrosio:1998yj},
while the ratio predicted with a unit form factor
is 0.21~\cite{bib:triangle}.

Within the VMD model, the value of the parameter $|a_S|$
can be obtained from the measurement of $B(K_S\rightarrow\pi^0\mu^+\mu^-)$
using Eq.~\ref{eq:as_vmd_mumu}:
\[  |a_S|_{\mu\mu} = 1.54_{-0.32}^{+0.40} \pm 0.06. \]
This value is in agreement with the same quantity extracted
from the study of $K_S\rightarrow\pi^0e^+e^-$
in Ref.~\cite{bib:piee}:
\[  |a_S|_{ee} = 1.06_{-0.21}^{+0.26} \pm 0.07.  \]
The ratio  $B(K_S \rightarrow \pi^0 \mu^+ \mu^-) /
B(K_S \rightarrow \pi^0 e^+ e^-)$ is found to be $0.49_{-0.29}^{+0.35}\pm 0.08$, 
in reasonable agreement with the VMD model prediction of 0.23.

\begin{figure}[hbtp]
  \vspace{9pt}
  \centerline{\hbox{ \hspace{0.0in}
    \epsfxsize=3.0in
    \epsffile{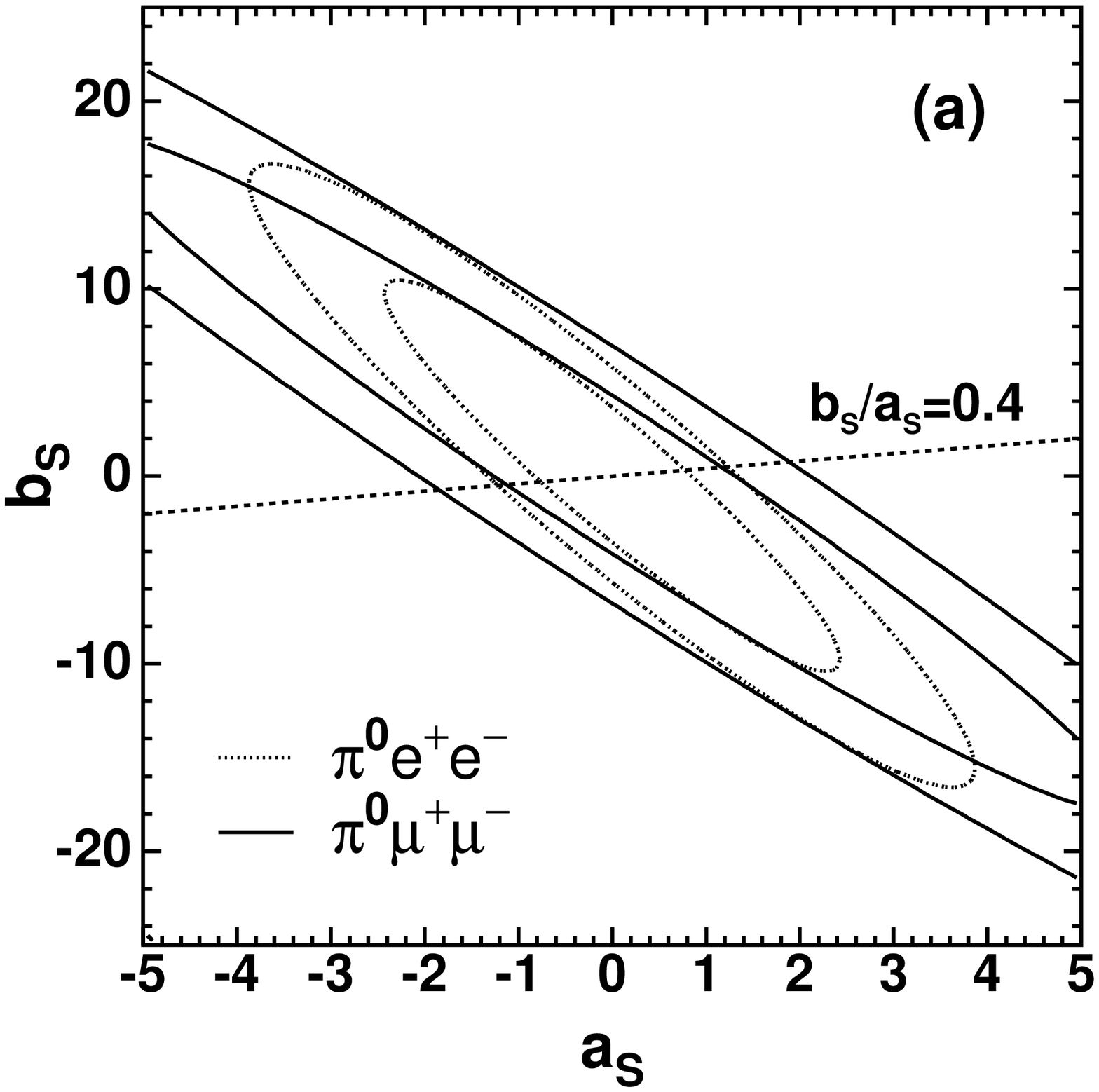}
    \hspace{0.25in}
    \epsfxsize=3.0in
    \epsffile{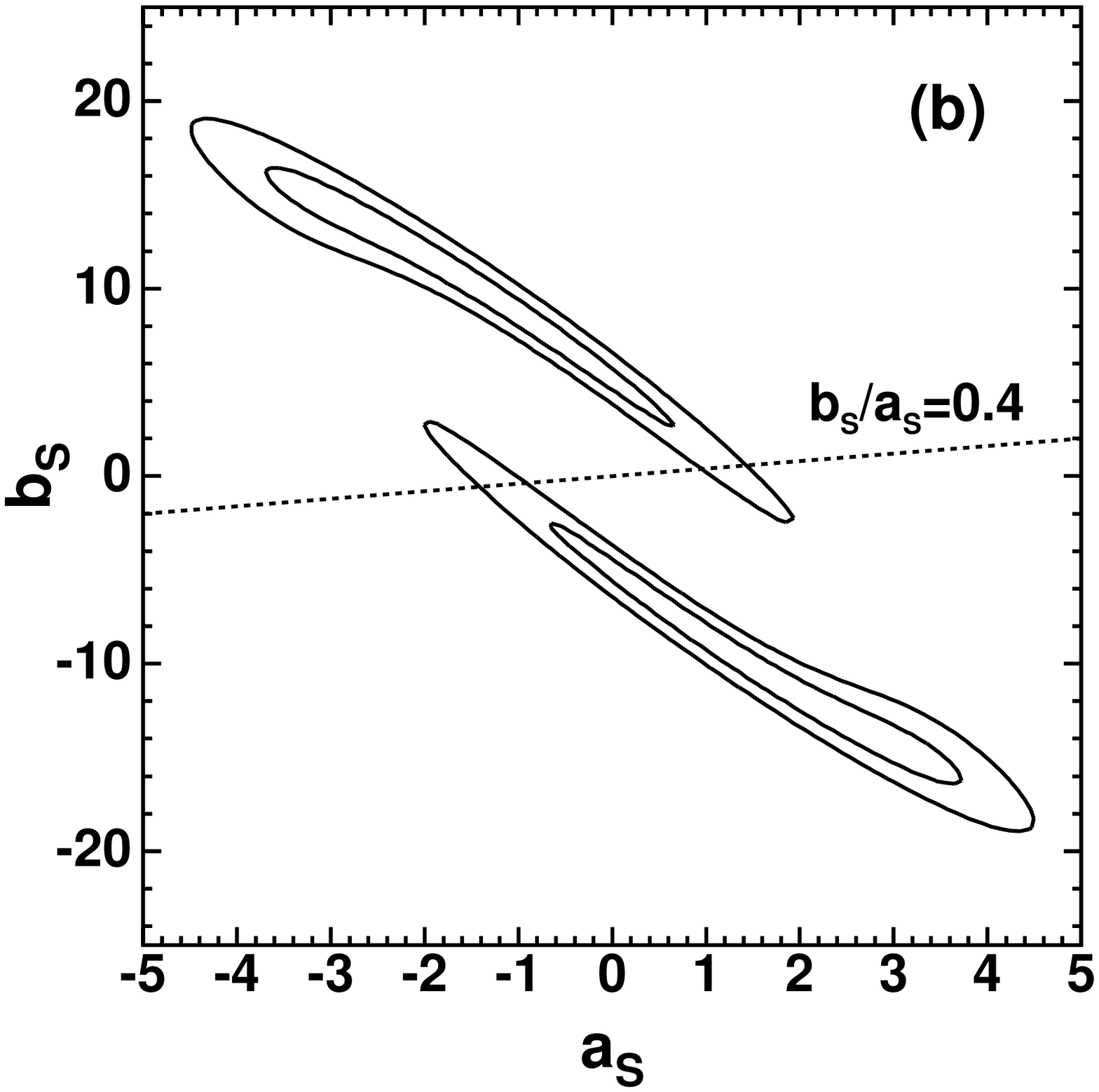}
    }
    }
\caption{(a) Allowed regions of $a_S$ and $b_S$ determined from the observed
number of $\kspimumu$ and $\kspiee$ events separately.
The region between the inner and outer solid (dashed) elliptical contours
is the allowed region for $\kspimumu$ ($\kspiee$) at 68\% CL.
(b) Allowed regions of $a_S$ and $b_S$
for the $\kspimumu$ and $\kspiee$ channels combined.
The inner (outer) contour of each pair delimits the $1\sigma$ ($2\sigma$)
allowed region from the combined log-likelihood.
The dashed straight line in both plots
corresponds to $b_S=0.4 a_S$, as predicted by the VMD model.
  \label{fig:ab} 
}
\end{figure}

The $a_S$ and $b_S$ parameters can be obtained from a combined analysis of 
$B(K_S \rightarrow \pi^0 e^+ e^-)$ and $B(K_S \rightarrow \pi^0 \mu^+ \mu^-)$
using Eqs.~\ref{eq:as_ee} and~\ref{eq:as_mumu}.
The 68\% confidence level contours in the $(a_S,b_S)$ plane derived from 
the measured decay rates are shown in Fig.~\ref{fig:ab}(a).
The VMD prediction $b_S/a_S=0.4$ falls within both sets of contours.

The constraints on $a_S$ and $b_S$
from the two branching ratio measurements 
have been combined using a maximum likelihood method,
as shown in Fig.~\ref{fig:ab}(b).
Two regions in the $(a_S,b_S)$ plane are preferred, namely
\begin{eqnarray}
a_S&=&-1.6_{-1.8}^{+2.1},~~~  b_S=~~\,10.8_{-7.7}^{+5.4} \\
a_S&=&~~\,1.9_{-2.4}^{+1.6},~~~  b_S=-11.3_{-4.5}^{+8.8}. 
\end{eqnarray}
The limited statistics do not allow a significant determination of $b_S$
or an assessment of the linear dependence of the form factor on $z$.
Table~\ref{tab:acceptance_m} shows the acceptance
as a function of dilepton mass 
to facilitate the comparison of the two decay channels.

\subsection{CPV component of {\boldmath $K_L\rightarrow\pi^0\mu^+\mu^-$}}

The measured branching ratio for the decay $\kspimumu$
allows the CPV contribution to the branching ratio of the corresponding $K_L$ decay,
$\klpimumu$, to be predicted as a function of
${\rm Im}(\lambda_t)$ to within a sign ambiguity \cite{Isidori:2004}:
\begin{equation}
B(K_L\rightarrow\pi^0\mu^+\mu^-)_{\rm CPV}\times  10^{12}= C_{\rm MIX}
\pm C_{\rm INT}  \left( \frac{\displaystyle {\rm Im}(\lambda_t)}{\displaystyle 10^{-4}}
\right) + C_{\rm DIR}  \left( \frac{\displaystyle {\rm Im}(\lambda_t)}{\displaystyle 10^{-4}}
\right)^2 ~, \label{eq:cpvt}
\end{equation}
where
\begin{equation*}
\nonumber
 C_{\rm MIX} = 3.1 \times 10^{9}\, B(K_S\rightarrow\pi^0\mu^+\mu^-),~~
 C_{\rm INT} = 4.6 \times 10^{4}\, \sqrt{B(K_S\rightarrow\pi^0\mu^+\mu^-)},~~
 C_{\rm DIR} = 1.0.  \nonumber
\label{eq:cmix}
\end{equation*}
$C_{\rm INT}$ is the coefficient for the term due to  
the interference between  the  direct\,($C_{\rm DIR}$) and 
indirect\,($C_{\rm MIX}$) CPV components, and $\lambda_t = V_{td}V_{ts}^*$. 

\begin{figure}[hbtp]
 \vspace{9pt}
 \centerline{\hbox{ \hspace{0.0in}
   \epsfxsize=3.0in
   \epsffile{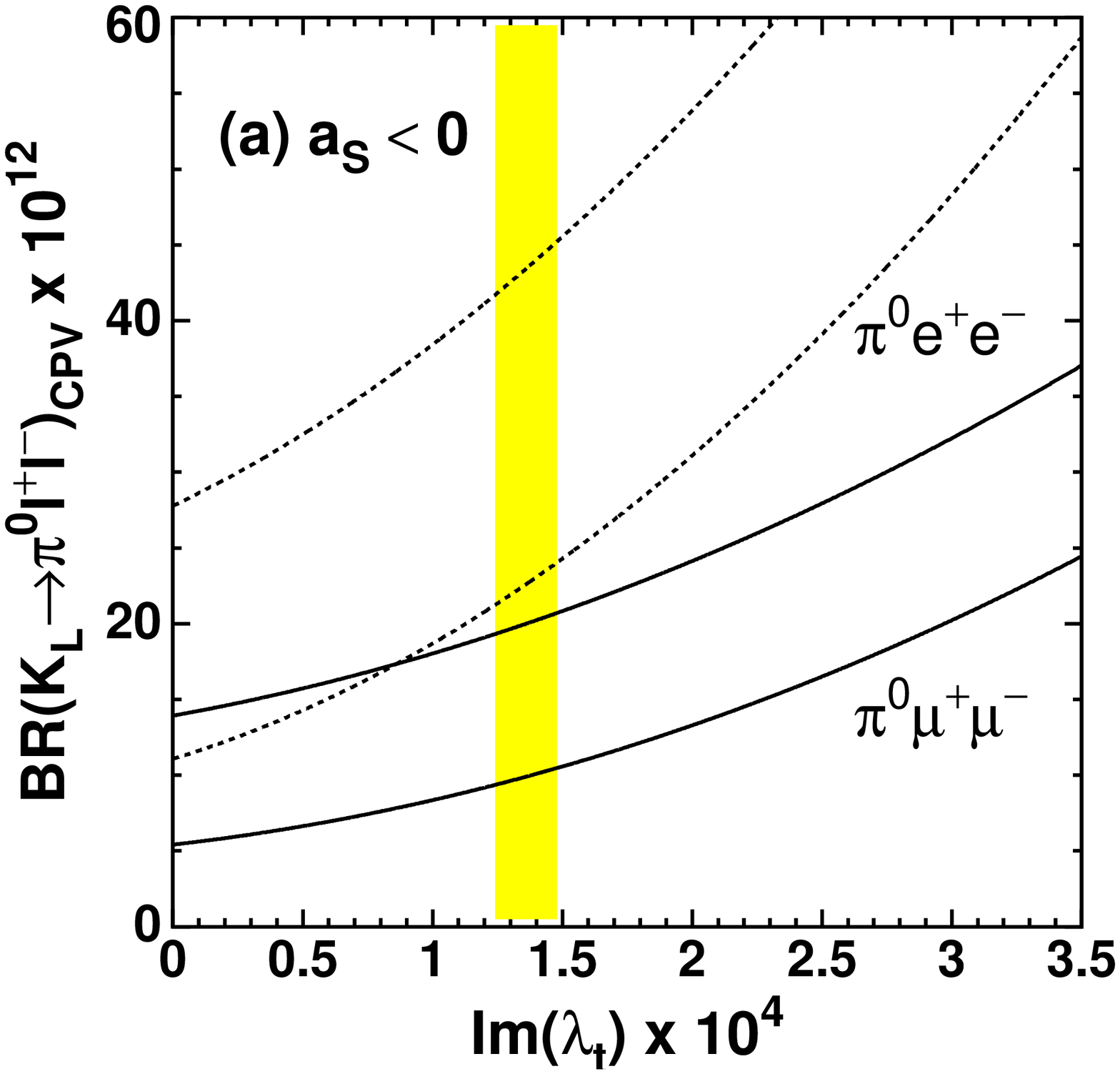}
   \hspace{0.25in}
   \epsfxsize=3.0in
   \epsffile{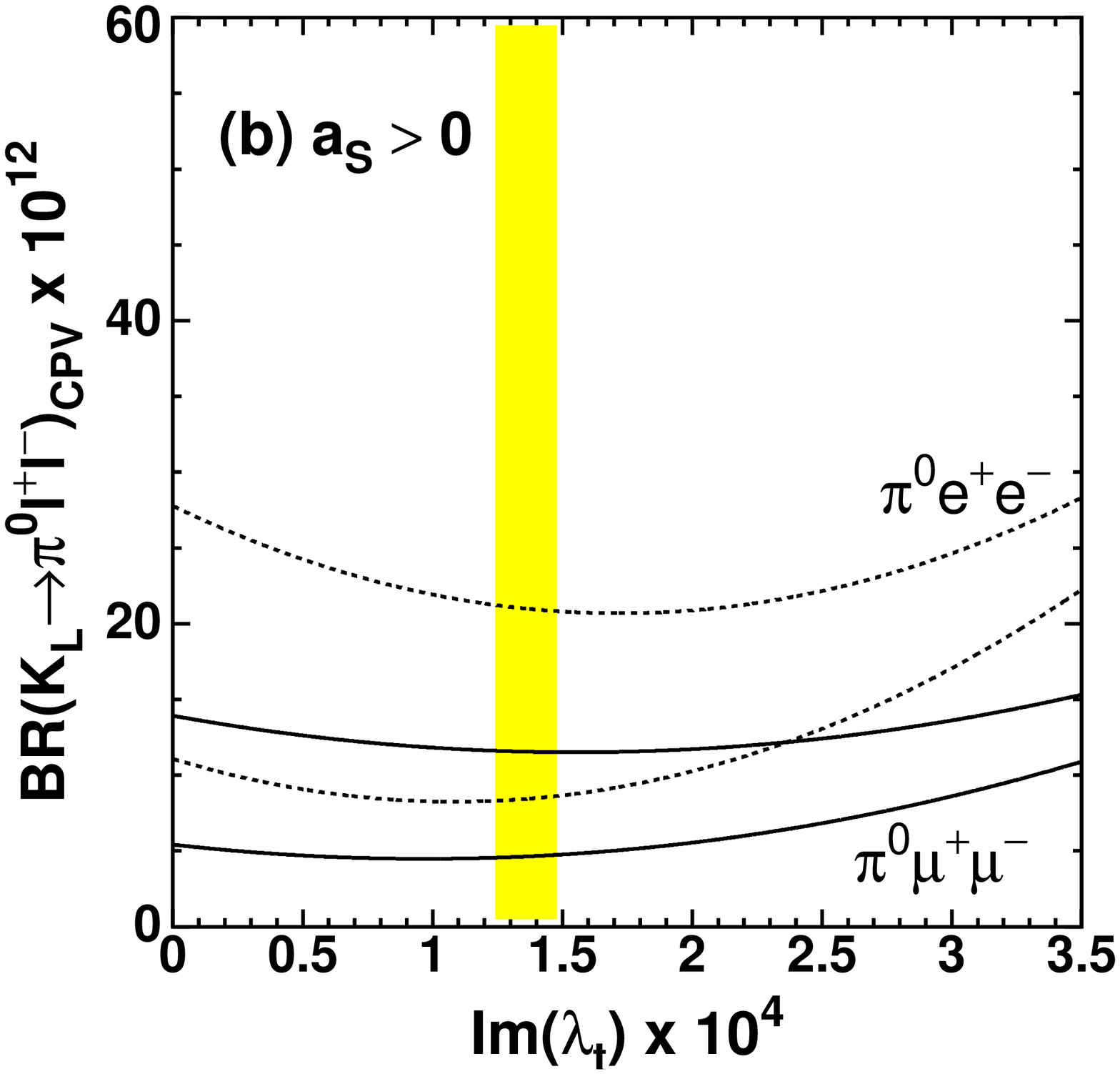}
   }
   }
\caption{Predicted CPV component of the $\klpimumu$ (solid curves)
and $\klpiee$ (dashed curves) branching ratios as a function of 
${\rm Im}(\lambda_t)$
assuming (a) $a_S<0$ and (b) $a_S>0$.
Each pair of curves delimits the allowed range derived from
the $\pm 1\sigma$ measured values of $|a_S|$.
The vertical shaded band shows the world average value of ${\rm Im}(\lambda_t)$
with its uncertainty~\cite{CKM}.
  \label{fig:KL} 
}
\end{figure}

The predicted dependence of $B(\klpimumu)_{\rm CPV}$
on ${\rm Im}(\lambda_t)$ is shown in Fig.~\ref{fig:KL}.
Taking the central value of the measured branching ratio 
$B(K_S\rightarrow\pi^0\mu^+\mu^-)$ and 
{${\rm Im}(\lambda_t)=1.36  \times 10^{-4}$}~\cite{CKM}
gives:
\begin{equation}
B(K_L \rightarrow \pi^0 \mu^+ \mu^-)_{CPV} \times 10^{12} \approx
8.8_{\rm mixing} \pm 3.3_{\rm interference} + 1.8_{\rm direct}.
\label{e:brklpi0mumu_cpv_cpts}
\end{equation}
The predicted dependence for the $\piee$ channel
is also shown in Fig.~\ref{fig:KL}.

The CP conserving (CPC) component of
$K_L \rightarrow \pi^0\ell^+\ell^-$ decays
can be constrained using measurements of the decay
$K_L\rightarrow\pi^0\gamma\gamma$~\cite{bib:piogg,bib:ktev_piogg}.
A recent analysis based on ChPT obtained the prediction
$(5.2\pm 1.6)\times 10^{-12}$~\cite{Isidori:2004}.

Combining the CPV and the CPC components,
the central value  for the total $\klpimumu$ branching ratio
is estimated to be $19 \times 10^{-12}$ or $13\times 10^{-12}$,
depending on the sign of the interference term between the direct and the 
indirect\,(mixing) amplitudes.
This estimate is consistent with the present experimental upper limit
on $B(\klpimumu)$ of $3.8 \times 10^{-10}$(90\% CL)~\cite{bib:ktev}.
The corresponding central value  for the total  $\klpiee$ branching ratio
was estimated to be $32\times 10^{-12}$ or $13\times 10^{-12}$~\cite{bib:piee}.

\section*{Acknowledgments}
It is a pleasure to thank the technical staff of the 
participating laboratories,
universities and affiliated computing centres for their efforts in the 
construction of the NA48 apparatus, in the operation of the experiment, and in 
the processing of the data.
We would also like to thank 
Gino Isidori 
and 
John Ellis 
for helpful discussions.

\newpage

\begin{table}
\begin{center}
\begin{tabular}{|c|c|c|c|}
\hline
\hline
 Event &  $K_S$ momentum (GeV) & $\tau / \tau_{S}$ & $m_{\ell\ell}$ (GeV)  \\
\hline
\hline
\multicolumn{4}{|c|}{$K_S\rightarrow \pi^0\mu^+\mu^-$}\\
\hline
1 &  93.1 & 0.48 & 0.241 \\
2 & 117.9 & 0.71 & 0.220 \\
3 & 147.8 & 1.74 & 0.226 \\
4 &  95.3 & 2.49 & 0.241 \\
5 & 112.7 & 0.64 & 0.262 \\
6 & 101.0 & 0.96 & 0.226 \\
\hline
\multicolumn{4}{|c|}{$K_S\rightarrow \pi^0 e^+e^-$}\\
\hline
1 & 84.6  & 0.74 & 0.291   \\
2 & 128.2 & 0.50 & 0.267   \\
3 & 114.1 & 1.02 & 0.173   \\
4 & 83.9  & 2.09 & 0.272   \\
5 & 130.8 & 1.46 & 0.303   \\
6 & 121.2 & 1.49 & 0.298   \\
7 & 94.2  & 1.64 & 0.253   \\
\hline
\hline
\end{tabular}
\hspace*{0.5cm}
\caption{ 
Kinematics of the six   events
found in the signal region for $K_S\rightarrow \pi^0\mu^+\mu^-$ and 
of the seven events found for $K_S\rightarrow \pi^0 e^+e^-$~\cite{bib:piee}.
\label{tab:events}}
\end{center}
\end{table}

\begin{table}
\begin{center}
\begin{tabular}{|l|c|}
\hline
\hline
Background Source     &  Events   \\
\hline
\hline
\klthreepich &  $0^{+0.02}_{-0.00}$  \\
\hline
\klmumugg    & $0.04^{+0.04}_{-0.03}$ \\
\hline
Accidentals & $0.18^{+0.18}_{-0.11}$   \\
\hline\hline
Total background &  $0.22^{+0.18}_{-0.11}$ \\
\hline
\hline
\end{tabular}
\caption{ Summary of the background estimate in the signal region.
\label{tab:back}}
\end{center}
\end{table}

\begin{table}
\begin{center}
\begin{tabular}{|c|c|}
\hline
\hline
  Total $K_S$ Flux &  $(2.50\pm 0.08)\times 10^{10}$ \\
\hline
Acceptance &  $0.081\pm 0.002\pm 0.004$  \\
\hline
Background & $0.22^{+0.18}_{-0.11}$\\
\hline
Events&  $6$ \\
\hline
\hline
\end{tabular}
\caption{ Summary of information used to extract the $\kspimumu$ branching ratio.
\label{tab:br}}
\end{center}
\end{table}

\begin{table}
\begin{center}
\begin{tabular}{|c|c||c|c|}
\hline
\hline
\multicolumn{2}{|c||}{$K_S\rightarrow \pi^0 \mu^+\mu^-$} &
\multicolumn{2}{c|}{$K_S\rightarrow \pi^0 e^+e^-$}\\
\hline
\hline
   $m_{\mu\mu}$ (GeV) & Acceptance & $m_{ee}$ (GeV) & Acceptance  \\
\hline
0.21-0.23 & $0.075^{+0.005}_{-0.008}$  &
0.000-0.0365 & $0.0809\pm 0.0015$  \\
0.23-0.25 & $0.072^{+0.005}_{-0.006}$  &
0.0365-0.073 & $0.0877\pm 0.0009$  \\
0.25-0.27 & $0.088^{+0.002}_{-0.003}$  &
0.073-0.1095 & $0.0910\pm 0.0007$  \\
0.27-0.29 & $0.083^{+0.003}_{-0.003}$  &
0.1095-0.146 & $0.0872\pm 0.0007 $  \\
0.29-0.31 & $0.084^{+0.004}_{-0.006}$  &
0.146-0.1825 & $0.0785\pm 0.0006$  \\
0.31-0.33 & $0.097^{+0.003}_{-0.014}$  &
0.1825-0.219 & $0.0755\pm 0.0006$  \\
0.33-0.35 & $0.079^{+0.003}_{-0.012}$  &
0.219-0.2555 & $0.0703\pm 0.0007$  \\
0.35-0.37 & $0.017^{+0.002}_{-0.002}$  &
0.2555-0.292 & $0.0664\pm 0.0008 $  \\
          &                            &
0.292-0.3285 & $0.0560\pm 0.0010$  \\
          &                            &
0.3285-0.365 & $0.0430\pm 0.0020$  \\
\hline
\hline
\end{tabular}
\caption{
Overall acceptance as a function of the dilepton mass for 
$\kspimumu$ and $\kspiee$~\cite{bib:piee}.
\label{tab:acceptance_m}}
\end{center}
\end{table}

\newpage

\end{document}